\documentclass[journal=jcisd8,manuscript=article,layout=twocolumn]{achemso}

\usepackage{multirow}
\usepackage{multicol}
\usepackage{hyperref}
\usepackage{array}
\usepackage{float}
\usepackage{chemformula}
\usepackage{siunitx}
\usepackage{microtype}
\usepackage{tabularx}
\usepackage{makecell}
\usepackage{booktabs} 
\usepackage{tabularx}
\usepackage{graphicx}
\usepackage[T1]{fontenc}
\usepackage{physics}
\usepackage{xurl}

\newcommand{\gnndMainTitle}{A Lightweight Universal Machine-Learning Interatomic Potential via Knowledge Distillation for Scalable Atomistic Simulations}
\newcommand{\gnnmlips}{MLIPs}
\newcommand{\gnnomni}{7net-Omni}
\newcommand{\gnnnano}{7net-Nano}
\newcommand{\gnnzero}{7net-0}

\author{Sangmin Oh}
\author{Jinmu You}
\author{Jaesun Kim}
\author{Jiho Lee}
\author{Hyungmin An}

\author{Seungwu Han}

\affiliation[SNU MSE]{Department of Materials Science and Engineering, Seoul National University, Seoul 08826, Korea}

\alsoaffiliation[KIAS]{AI center, Korea Institute for Advanced Study, Seoul 02455, Korea}
\alsoaffiliation[SNU RIAM]{Research Institute of Advanced Materials, Seoul National University, Seoul 08826, Korea}

\author{Youngho Kang}
\email{youngho84@inu.ac.kr}
\affiliation[INU]{Department of Materials Science and Engineering, Incheon National University, Incheon 22012, Korea}

\title[maintitle]
  {\gnndMainTitle}
\keywords{machine learning interatomic potential, graph neural network potential,  
molecular dynamics, distillation}

\begin{document}

\begin{abstract}
We introduce a lightweight universal machine-learning interatomic potential (uMLIP), SevenNet-Nano, based on the graph neural network architecture SevenNet and enabled by a knowledge-distillation framework. The model inherits the broad generalization capability of a large multi-task foundation model, SevenNet-Omni, trained on diverse materials datasets across chemical, configurational, and computational spaces. By learning chemical representations from high-quality inference data generated by the teacher model within a unified computational framework, SevenNet-Nano achieves high accuracy and strong transferability despite its compact architecture. The model also accurately captures a wide range of interatomic interactions, enabling reliable simulations under both equilibrium and extreme conditions, including plasma etching of \ch{SiO2}. Comprehensive benchmarks on static and dynamical properties—such as Li-ion diffusion and liquid densities—demonstrate its broad applicability with minimal fine-tuning. Importantly, SevenNet-Nano significantly reduces computational cost, achieving over an order-of-magnitude speedup and enabling large-scale atomistic simulations involving thousands of atoms.
\end{abstract}






\maketitle
\section{Introduction}

Machine-learning interatomic potentials (MLIPs) have emerged as a transformative approach for atomistic simulations by bridging the gap between the high accuracy of quantum mechanical methods and the efficiency of classical force fields\cite{mlip_behler2016perspective,mlip_deringer2019machine}. Early developments of MLIPs focused on descriptor-based models tailored to specific materials systems, where local atomic environments are encoded using handcrafted descriptors such as atom-centered symmetry functions\cite{bpnn_behler2007generalized}, moment tensor basis \cite{mtp_shapeev2016moment,mtp_novikov2021mlip}, or smooth overlap of atomic positions (SOAP)\cite{soap_de2016comparing}. Combined with regression models including neural networks and Gaussian processes, these approaches achieved near–density functional theory (DFT) accuracy for energies and forces in targeted systems. However, their applicability is inherently limited by the system-specific nature of both descriptors and training datasets, requiring substantial retraining to extend across different chemistries, phases, or defect configurations.

To overcome these limitations, recent efforts have shifted toward universal pretrained MLIPs (uMLIPs), such as Orb \cite{orb_rhodes2025orb}, UMA \cite{uma_wood2025family}, DPA \cite{dpa_zhang2025graph}, eSEN \cite{esen_fu2025learning}, GRACE \cite{b_gb_lysogorskiy2026graph}, MatterSim \cite{mattersim_yang2024mattersim}, MACE-MH \cite{macemh_batatia2025cross} and SevenNet-Omni (hereafter {\gnnomni}) \cite{omni_kim2025optimizing}, trained on large and diverse datasets spanning broad chemical and configurational spaces. Enabled by equivariant graph neural networks (GNNs), these models learn representations of atomic interactions directly from DFT data, eliminating the need for handcrafted descriptors and significantly improving transferability. As a result, uMLIPs have demonstrated strong generalization across molecules, crystalline solids, defects, and amorphous structures, positioning them as emerging foundation models for atomistic simulations.

Despite their success, achieving such universality typically requires large model capacity, including deep architectures, high-order equivariant features, and extended interaction layers~\cite{scale_qu2024importance,scale_liao2023equiformerv2,scale_leimeroth2025machine,esen_fu2025learning}. These design choices may result in substantial computational overhead in both memory usage and inference time, limiting the practical applicability of state-of-the-art uMLIPs in large-scale molecular dynamics (MD) simulations. While reducing model size improves efficiency, directly training lightweight models on heterogeneous datasets leads to degraded accuracy and reduced transferability. This challenge highlights a fundamental trade-off between model efficiency and generalization capability in uMLIPs. Large models excel at capturing complex potential energy surfaces (PESs) across diverse systems but are computationally expensive, whereas small models are efficient but struggle to maintain comparable accuracy when trained from scratch. Bridging this gap is essential for enabling scalable atomistic simulations \cite{fast_xie2023ultra,fast_kong2026scalable}.

In this context, knowledge distillation via teacher–student learning provides a promising solution. In this framework, a large universal model serves as a teacher, transferring its learned representation of the PES to a lightweight student model through supervised signals such as energies and forces~\cite{mixedloss_taniguchi2025knowledge}. Previous studies have also incorporated additional supervisory signals, including atomic energies~\cite{ae_jung2025atomic, ae_Matin_2025}, energy Hessian matrices~\cite{hessian_Sanjeev_2025}, and intermediate architectural information such as node and edge features~\cite{feature_ekstrom2023accelerating}. These works have primarily focused on demonstrating the effectiveness of distillation methods for specific target tasks. On the other hand, a previous study proposed a framework for training a target-specific MLIP, termed LightPFP~\cite{lightpfp_li2025}, by leveraging a pretrained student model to achieve a better balance between computational efficiency and accuracy. In LightPFP, a lightweight student model based on the moment tensor potential (MTP)~\cite{mtp_shapeev2016moment, mtp_novikov2021mlip} is first pretrained and subsequently fine-tuned using data generated by the teacher Preferred Potential (PFP)~\cite{pfp_takamoto2022towards} for a specific task. While this approach demonstrates the feasibility of distilling knowledge into a compact model and its applicability to diverse systems, fine-tuning of the pretrained student model remains essential for achieving reliable accuracy in materials simulations. However, such fine-tuning may involve complex error-estimation and sampling procedures, even requiring multiple iterations to achieve satisfactory accuracy. This can offset the practical advantages of a pretrained student model, thereby motivating the development of cost-effective approaches that minimize the need for system-specific retraining.

\begin{figure*}[!t] 
\centering
\includegraphics[width=1.0\textwidth]{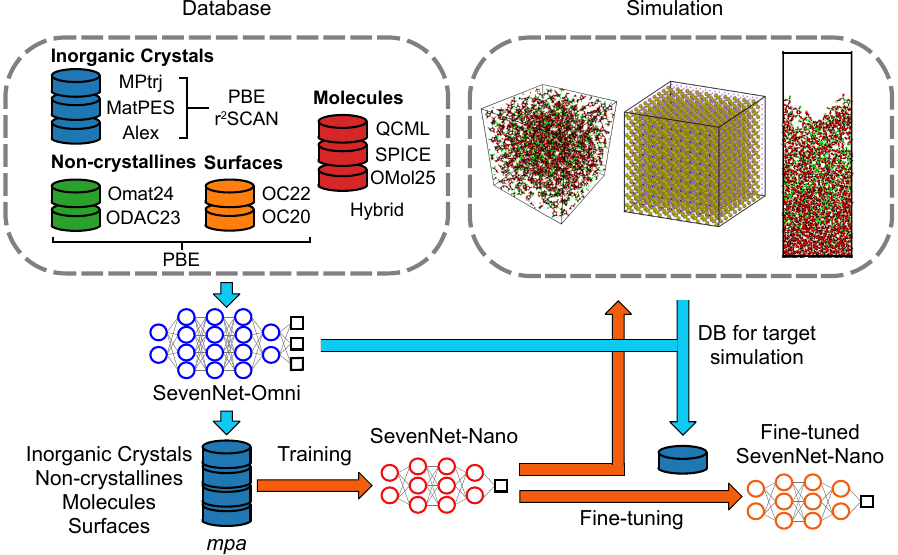}
\caption{Schematic illustration of the training of {\gnnnano} via knowledge distillation from {\gnnomni}, along with its fine-tuning procedure.}
\label{fig:full_schema} 
\end{figure*}

In this work, we introduce a lightweight universal MLIP using the GNN-based SevenNet architecture, denoted as SevenNet-Nano (hereafter {\gnnnano}), enabled by a distillation framework that transfers knowledge from a large foundation model to an efficient yet general-purpose potential suitable for scalable atomistic simulations. The key idea is to leverage inference data generated by a multi-task teacher model, {\gnnomni}, which is trained on diverse materials datasets spanning wide chemical and configurational spaces, as well as multiple exchange–correlation functionals. Owing to its multi-task nature, {\gnnomni} can generate high-accuracy inference data within a specific computational setting (i.e., a single-fidelity channel) across vast training domains. By training on this dataset, the student model inherits the generalization capability of the teacher through knowledge distillation (\autoref{fig:full_schema}). As a result, the student model achieves improved accuracy and transferability compared to uMLIP models trained directly on limited datasets, even with a lightweight architecture that would otherwise be insufficient to capture complex representations across multiple domains from scratch. In particular, {\gnnnano} exhibits high accuracy for sub-angstrom short-range interatomic interactions, enabling reliable materials simulations under extreme conditions, including semiconductor processes such as plasma etching. At the same time, {\gnnnano} significantly reduces computational cost, enabling large-scale simulations involving thousands of atoms and achieving speedups of over an order of magnitude compared to its teacher model. Through comprehensive benchmarks on both static and dynamical properties—including Li diffusion, liquid densities, and \ch{SiO2} etching—we demonstrate the broad applicability of the model, requiring only minimal fine-tuning based on data from the teacher model when necessary.

\section{Computational Methods}

\subsection{Training of {\gnnnano}}

We adopt a single-task SevenNet architecture to train {\gnnnano}, using a maximum spherical harmonic order of $l_\mathrm{max}=2$ and a dimension of 32 for all equivariant node features, along with three convolution layers. This architecture comprises a total of 105~k parameters, representing a substantial reduction compared to the 26~M parameters of the {\gnnomni} teacher model. The teacher model employs $l_\mathrm{max}=3$ and five convolution layers, with node feature dimensions of 128, 64, and 32 for $l=0$, $l=1$, and $l>1$, respectively.
 
For training, we define the loss function including atomic energies:
\begin{equation}
\begin{aligned}
\mathcal{L} &= \frac{\lambda_{\mathrm{E}}}{M} \sum_{i=1}^{M} \frac{|\hat{E}_i - E_i|}{N_i} \\
&\quad + \frac{\lambda_{\mathrm{F}}}{M \sum_{i=1}^{M} N_i} \sum_{i=1}^{M} \sum_{j=1}^{N_i} \sqrt{\sum_{k=1}^{3} |\hat{F}_{i,j,k} - F_{i,j,k}|^2} \\
&\quad + \frac{\lambda_{\mathrm{S}}}{M} \sum_{i=1}^{M} \sqrt{\sum_{j=1}^{3} \sum_{k=1}^{3} |\hat{S}_{i,j,k} - S_{i,j,k}|^2} \\
&\quad + \frac{\lambda_{\mathrm{AE}}}{M} \sum_{i=1}^{M} \sum_{j=1}^{N_i} \frac{|\hat{AE}_{i,j} - AE_{i,j}|}{N_i} \\
\end{aligned}
\label{eq:loss}
\end{equation}
where $M$ denotes the number of structures in a batch and $N_i$ is the number of atoms in structure $i$.
$E_i$, $F_{i,j,k}$, $S_{i,j,k}$, and $AE_{i,j}$ denote the teacher-model total energy, force (the $k$-th Cartesian component acting on atom $j$), stress (the $(j,k)$ component of the symmetric Cauchy stress tensor), and atomic energy of atom $j$ for structure $i$, respectively. The corresponding inference data from {\gnnmlips} are represented by $\hat{E}_i$, $\hat{F}_{i,j,k}$, $\hat{S}_{i,j,k}$, and $\hat{AE}_{i,j}$. The loss weights are set to $\lambda_{\mathrm{E}} = 1.0$, $\lambda_{\mathrm{F}} = 1.0$, $\lambda_{\mathrm{S}} = 2 \times 10^{-4}$, and $\lambda_{\mathrm{AE}} = 0.5$. The corresponding units of the quantities are eV for energies, eV/\AA\ for forces, and kbar for stresses. 

We examine four different cutoff radii ($r_c$) of 4.5, 5.0, 5.5, and 6.0~\AA, which determine the interaction range for edges connecting atoms, while maintaining an identical neural-network architecture. Below, the corresponding models are denoted as {\gnnnano}-4.5, {\gnnnano}-5.0, {\gnnnano}-5.5, and {\gnnnano}-6.0, respectively. The model weights are randomly initialized, except for the scale and shift parameters, which are extracted from the mpa channel of {\gnnomni}. These parameters are introduced to improve learning efficiency by preventing significant differences in the reference energies across elements during training. The shift parameters are trainable during training, whereas the scale parameters are kept fixed. Each {\gnnnano} model with a different cutoff radius is trained for two epochs using the AdamW optimizer~\cite{adamw_loshchilov2017decoupled}.

To construct the training dataset, we consider multiple configuration databases that were originally used to train {\gnnomni} \cite{omni_kim2025optimizing}. These include datasets for inorganic crystals (MPtrj \cite{mptrj_deng2023chgnet}, MPtrj-r2SCAN \cite{mpr2scan_kingsbury2022performance,mpr2scan_merchant2023scaling}, Alex \cite{alex_schmidt2023machine,alex_schmidt2024improving}, OMat24 \cite{omat24_barroso2024open}, MatPES and MatPES-r2SCAN \cite{matpes_kaplan2025foundational}), molecular systems (OMol25 \cite{omol25_levine2025open}, SPICE \cite{spice_eastman2023}, and QCML \cite{qcml_ganscha2025}), surfaces (OC20 \cite{oc20_chanussot2021open} and OC22 \cite{oc22_tran2023open}), metal-organic frameworks (ODAC23 \cite{odac23_sriram2024open}), and multi-domains (MAD \cite{mad_mazitov2025massive}). The training data is generated by evaluating their energies, forces, and stresses using the mpa channel of {\gnnomni}, corresponding to the Perdew–Burke–Ernzerhof (PBE) exchange–correlation functional used in DFT. In addition, to enhance fidelity to the teacher model, the corresponding atomic energies are added to the training set. To balance the representation across different training domains, oversampling is applied to selected datasets of configurations~\cite{omni_kim2025optimizing}. We prepare the database using the torch{\_}sim package \cite{torchsim_cohen2025}. The dataset is split into training and validation subsets using the same procedure as for {\gnnomni}.

\subsection{MLIP calculations}

SevenNet MLIP calculations are performed using the LAMMPS package \cite{lammps_thompson2022}, with FlashTP \cite{leeflashtp} for acceleration, unless otherwise stated. We calculate the Li-ion diffusivity of solid-state electrolytes, including eight argyrodite compounds of the form Li$_{24+x}$M$_4$S$_{20}$X$_4$, where M denotes group IV and V cations and X represents halide anions (see specific compositions in Table~S1), as well as non-argyrodite materials. The latter include oxides, such as \ch{Li7La3Zr2O12} (LLZO) and Li$_{1.33}$Ti$_{1.67}$Al$_{0.33}$(PO$_4$)$_3$ (LATP) \cite{sse_ox_he2018statistical,sse_ox_murugan2007fast}; sulfides, including \ch{Li10GeP2S12} (LGPS) and \ch{Li7P3S11} (LPS) \cite{sse_sul_kamaya2011lithium,sse_sul_seino2014sulphide}; halides, such as \ch{Li3YCl6} (LYC) and \ch{Li3YBr6} (LYB) \cite{sse_hal_asano2018solid}; and nitrides, including \ch{Li3N} and \ch{Li9S3N} (LSN) \cite{sse_nit_li2010li+,sse_nit_miara2015li}. Li diffusion in these systems was systematically examined using MLIP approaches in a previous study \cite{ft_kim2025efficient}. The initial structures are obtained by relaxing the atomic models used in the previous study to evaluate Li diffusion~\cite{ft_kim2025efficient} employing the Atomic Simulation Environment (ASE) package \cite{ase_hjorth2017atomic}. The models are designed to have cell dimensions of approximately 10~\AA\ to minimize self-interactions between neighboring periodic images. A force tolerance of \SI{0.04}{eV/\angstrom} is applied during structural optimization. For systems known to exhibit cubic symmetry and site disorder, such as argyrodites and LLZO, we constrain the cell to remain cubic during structural relaxation. Subsequently, we perform NVT MD simulations using a Nosé–Hoover thermostat with a timestep of \SI{2}{\fs} for \SI{100}{\ps}. Simulations are carried out at three temperatures of 800, 1000, and 1200~K. For each temperature, five independent runs with randomly assigned initial velocities are executed. The mean-square displacement (MSD) of Li ions is evaluated using the time-averaged approach \cite{sse_ox_he2018statistical}, and the Li-ion diffusivity for each run is obtained from the slope of the MSD–time curve in the linear regime, divided by the dimensional factor of 6.

To compute the densities of twenty solvents for Li-ion liquid electrolytes listed in \autoref{table:solvent_abbrev}, we follow the simulation procedure reported in a previous study \cite{ft_Li_ju2025application}. The initial configurations are generated to contain approximately 1,000 atoms within a cubic cell using MolView~\cite{molview_bergwerf} and PACKMOL~\cite{packmole_hurle1982self}. Considering the van der Waals volumes of the molecules, the initial cell dimensions and molecular positions are chosen such that the molecules can interact within the cutoff radius of the MLIPs while avoiding excessively short intermolecular distances. The atomic positions are first relaxed with a coarse force tolerance of \SI{2.0}{eV/\angstrom}. Subsequently, NPT MD simulations are performed at ambient pressure using a Nosé–Hoover thermostat. During the pre-equilibration stage, MD simulations are carried out for \SI{0.4}{\ns} using a relatively large timestep of \SI{2}{\fs} and an increased hydrogen mass of \SI{3}{\text{a.u.}} In the subsequent 0.4-ns equilibration run, the timestep is reduced to \SI{1}{\fs}, and the hydrogen mass is restored to its original value of \SI{1}{\text{a.u.}} The temperature is set to \SI{298}{\K}, except for EC (\SI{313}{\K}), FEC (\SI{313}{\K}), DME (\SI{293}{\K}), and DEE (\SI{293}{\K}). Finally, the equilibrium density is obtained by averaging the instantaneous density sampled every \SI{1}{\ps} over the final \SI{0.2}{\ns} of the simulation. The pre-equilibration time used in this work (\SI{0.4}{\ns}) is shorter than that used in the previous study (\SI{1}{\ns})~\cite{ft_Li_ju2025application}, but it has a negligible effect on the calculated densities, with differences within 1{\%}.

\begin{table}[!t]
    \centering
    \caption{Classifications, IUPAC names, and abbreviations of the Li-electrolyte solvents used in this study.}
    \small
    \setlength{\tabcolsep}{3pt}
    \begin{tabular}{c p{0.45\columnwidth} c}
        \toprule
        \textbf{Type} & \textbf{IUPAC name} & \textbf{Abbreviation} \\
        \midrule
        \multirow{7}{*}{\makecell{cyclic \\ carbonate}} 
        & Ethylene carbonate & EC \\
        & Fluoroethylene carbonate & FEC \\
        & Propylene carbonate & PC \\
        & Vinylene carbonate & VC \\
        & Difluoroethylene carbonate & DFEC \\
        & \textit{cis}--Difluoroethylene carbonate & \textit{cis}--DFEC \\
        & \textit{trans}--Difluoroethylene carbonate & \textit{trans}--DFEC \\
        \midrule
        \multirow{6}{*}{\makecell{linear \\ carbonate}} 
        & Diethyl carbonate & DEC \\
        & Dimethyl carbonate & DMC \\
        & Ethyl methyl carbonate & EMC \\
        & Fluoromethyl methyl carbonate & MFDMC \\
        & Bis(fluoromethyl) carbonate & DFDMC \\
        & Difluoromethyl fluoromethyl carbonate & TFDMC \\
        \midrule
        \multirow{2}{*}{ether} 
        & 1,2-Dimethoxyethane & DME \\
        & 1,2-Diethoxyethane & DEE \\
        \midrule
        \multirow{5}{*}{ester} 
        & $\gamma$-Butyrolactone & GBL \\
        & Ethyl acetate & EA \\
        & Fluoroethyl acetate & FEA \\
        & Ethyl 2-fluoroacetate & EFA \\
        & Ethyl propionate & EP \\
        \bottomrule
    \end{tabular}
\label{table:solvent_abbrev}
\end{table}

Plasma etching of \ch{SiO2} using \ch{CF2} and \ch{CF3} ions as reactive species is simulated using MLIPs. Amorphous \ch{SiO2} bulk structures are generated by randomly placing atoms within a cubic cell, followed by NVT MD simulations consisting of a pre-melting stage (5000 K, 2 ps) and a melt–quench–anneal process: melting (4000 K, 20 ps), quenching (from 4000 K to 300 K at a rate of –100 K/ps), and annealing (500 K, 15 ps). The density of the amorphous structure is set to 2.34 g/cm$^3$, corresponding to the experimental value~\cite{vsimurka2018mechanical,dehghani2021effect}. A \ch{SiO2} slab model representing the substrate is then constructed by introducing a 20-nm vacuum along the $z$-axis. The area and thickness of the amorphous \ch{SiO2} substrate vary within 1–9 nm$^2$ and 4–7 nm, respectively. The bottom 4–6~\AA\ layer is kept fixed to prevent drift motion. We adopt the same etching simulation protocol as in a previous study~\cite{chfetch_An_2026}. Specifically, we focus on neutral \ch{CF2} and \ch{CF3} species, assuming that corresponding ions are neutralized near the substrate surface prior to impact via Auger neutralization, as commonly assumed in previous etching simulations~\cite{nion_tully1977neutralization,nion_pretzer1966ion}. The influence of ionic acceleration in the plasma phase is incorporated by varying the initial kinetic energy: neutral \ch{CF2} and \ch{CF3} molecules are injected at normal incidence onto the substrate with kinetic energies ranging from 20 to 1000 eV for a specified number of impacts (100–200 ions/\SI{}{\nm\squared}). The detailed simulation conditions depend on the supercell size, as shown in Table~S2. After each ion incidence, a 2-ps NVE MD simulation is performed to account for energy transfer, followed by NVT cooling with a Langevin thermostat until the slab temperature reaches 300 K~\cite{langevin_allen2017computer}. To prevent excessive atomic displacements during high-energy impacts, we adaptively adjust the timestep to restrict the maximum displacement to \SI{0.1}{\angstrom} per step, limit the kinetic energy transfer to \SI{1}{eV} per atom, and cap it with an initial timestep of \SI{1}{fs}. We remove volatile byproducts, if formed during etching simulations, including \ch{SiF2}, \ch{SiF4}, \ch{CF2}, \ch{CF4}, \ch{O2}, \ch{F2}, \ch{CO}, \ch{OF2}, \ch{COF2}, and \ch{CO2}~\cite{mogab1978plasma,ionbeam_toyoda2004beam}, which remain attached to the surface after each incidence of ions. To maintain sufficient substrate thickness during etching, the slab thickness is monitored every 10 ion impacts. If the thickness decreases by more than 1 nm from the initial value, the substrate is replenished to recover the original thickness by adding a portion of the bulk \ch{SiO2} to the bottom of the slab~\cite{chfetch_An_2026}. During the etching and liquid-electrolyte simulations, the Grimme-D3 van der Waals (vdW) correction is included to account for long-range dispersion interactions. 

\subsection{DFT calculations}

For the DFT calculations, we use the Vienna Ab initio Simulation Package (VASP) \cite{vasp_kresse1999ultrasoft}. The PBE exchange-correlation functional and projector-augmented-wave (PAW) pseudopotentials are employed. Throughout this study, a plane-wave cutoff energy of 520 eV is used. The pseudopotential version and cutoff energy are consistent with those used to generate the MPtrj dataset. Spin polarization is included. A $3\times3\times3$ $k$-point grid is used for 64-atom Si, SiC, and 99-atom SiO$_2$ simulations, while only the $\Gamma$-point is sampled for other systems. This $k$-point sampling is as dense as or denser than that suggested by the pymatgen code~\cite{pymatgen_ong2013python}, which is used for the MPtrj database.

\section{Results and discussion}
\subsection{Benchmarks of {\gnnnano}} 
\begin{figure*}[!t] 
\centering
\includegraphics[width=1.0\textwidth]{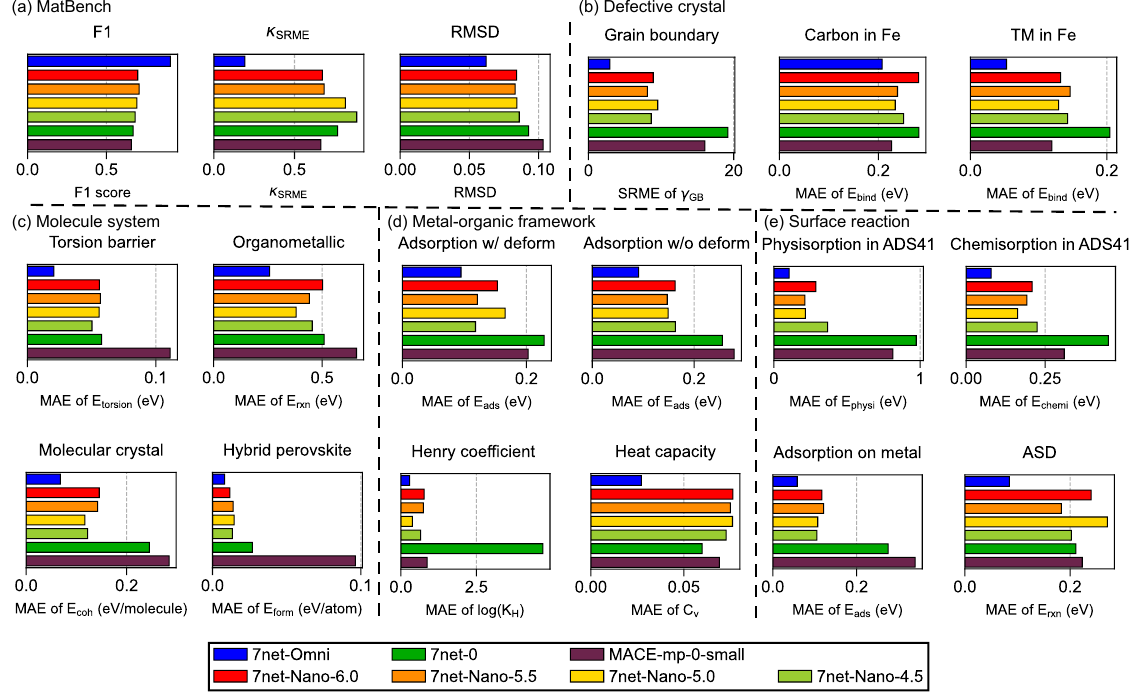}
\caption{Benchmark performance of {\gnnnano} across the standard tasks used for {\gnnomni} validation: (a) Matbench, (b) defective crystals, (c) molecular systems, (d) metal-organic frameworks, and (e) surface reactions. Metrics and error values of {\gnnomni} are adopted from Ref.~\cite{omni_kim2025optimizing}. TM and ASD denote transition metal and area-selective deposition, respectively.}
\label{fig:bench_nano} 
\end{figure*}

We first check the training quality by calculating the mean absolute errors (MAEs) of {\gnnnano} on the validation dataset—a subset of the {\gnnomni}-generated training data—after two training epochs (Figure~S1). The $r_c$-dependent MAE ranges for energy, atomic energy, force, and stress are 23–27 meV/atom, 0.046–0.050 eV, 0.078–0.083 eV/\AA, and 4.52–4.82 kbar, respectively. These values are reasonably low and comparable to those reported for other GNN-based MLIP models (except for the MAE for atomic energy, which is not available for such models), such as SevenNet-0 (hereafter {\gnnzero}) and MACE\cite{mace_batatia2025foundation}, indicating good training quality.

To evaluate the accuracy of the {\gnnnano} models across diverse chemical domains, we perform the same benchmark tests used to assess the teacher model {\gnnomni} (\autoref{fig:bench_nano}). The reference data are the corresponding PBE(+D3) results, except for the Henry coefficient, for which experimental data are used as the reference. The D3 dispersion correction is applied in all calculations except for the MatBench, defective crystal, and adsorption on metal from \cite{omni_kim2025optimizing} cases. For comparison, we also present results for {\gnnomni} and other single-task MLIPs, such as {\gnnzero} and MACE-mp-0-small. The latter two models are trained solely on the MPtrj inorganic crystal dataset. As such, comparison with these models helps us assess the domain coverage of the {\gnnnano} models. The sources of the reference DFT data are listed in Table~S3~\cite{omni_kim2025optimizing,matbench_riebesell2025framework,
b_gb_lysogorskiy2026graph,b_gb_zheng2020grain,
b_df_becquart2007atomistic, b_df_olsson2010ab,
b_tor_lahey2020benchmarking, b_tor_rai2022torsionnet,  b_om_dohm2018comprehensive, b_om_gusev2013assessing, b_mc_reilly2013understanding, b_mc_moellmann2014dft, b_mc_zhugayevych2023benchmark, b_pv_kim2017hybrid, b_asd_kim2026computational, b_ads41_mallikarjun2019adsorption, b_mof_krass2025mofsimbench, b_mof_moosavi2022data, b_mof_odac25_sriram2025open, b_mof_golddac_lim2025accelerating, b_mof_brabson2025comparing}. We begin by evaluating the F1 score, $\kappa_{\rm{SRME}}$, and RMSD reported from the MatBench Discovery benchmark (\autoref{fig:bench_nano}a), which serve as metrics to evaluate model performance on the inorganic crystals. The F1 score measures the accuracy of crystalline energy ranking, with a higher value indicating better performance. $\kappa_{\rm{SRME}}$ represents the symmetric relative mean error (SRME) of the lattice thermal conductivity ($\kappa$). RMSD quantifies structural deviations after relaxation relative to DFT. For all three benchmark metrics, {\gnnomni} achieves the best performance. The student model {\gnnnano} exhibits reasonable accuracy, albeit lower than that of {\gnnomni}, with performance comparable to that of {\gnnzero} and MACE-mp-0-small, both built upon the MPtrj dataset.

Beyond pure inorganic crystals, we evaluate the accuracy of the models for out-of-domain defective structures, including grain boundaries (GBs) and point defects (\autoref{fig:bench_nano}b). We compute the SMREs for 297 GB configurations across 58 metallic elements, following the benchmark protocol previously reported for {\gnnomni}. Notably, for GBs, {\gnnnano} outperforms {\gnnzero} and MACE-mp-0-small, benefiting from the broader domain coverage and improved representations inherited from the teacher model. For the binding energies of vacancy–solute complexes, including Fe vacancy–interstitial carbon and Fe vacancy–substitutional transition metals (TMs) in bcc iron, {\gnnnano} maintains accuracy comparable to that of other single-task models. 

For molecular systems, {\gnnnano} effectively captures the key representations of {\gnnomni}, yielding better performance than {\gnnzero} and MACE-mp-0-small (\autoref{fig:bench_nano}c). Specifically, {\gnnnano} achieves lower MAEs for small-molecule torsion barriers, which assess the ability to describe conformational PESs, across 188 molecules from two pharmaceutically relevant datasets, the Biaryl set and TorsionNet500. {\gnnnano} also yields lower MAEs for the reaction energies of organometallic systems ($E_{\mathrm{rxn}}$), evaluated over 53 reactions involving 97 organometallic complexes. For the cohesive energies of molecular crystals ($E_{\mathrm{coh}}$), defined as the energy required to separate a crystal into individual molecules, {\gnnnano} shows moderate improvement. Furthermore, for 100 organic–inorganic hybrid perovskites ABX$_3$, which are emerging materials for photovoltaics and optoelectronics~\cite{hybprv_egger2016hybrid,hybprv_moroni2024chiral,hybprv_berry2015hybrid}, {\gnnnano} predicts formation energies with reduced MAEs.

The higher accuracy of {\gnnnano} compared to {\gnnzero} and MACE-mp-0-small is also observed for metal–organic frameworks (MOFs). The \ch{H2O} and CO$_2$ adsorption energies ($E_{\mathrm{ads}}$) within MOFs, which have been extensively studied for environmental applications such as carbon capture~\cite{mof_ding2019carbon,mof_simmons2011carbon}, are evaluated using MLIPs (\autoref{fig:bench_nano}d). Regardless of whether deformation is allowed, the MAEs of the {\gnnnano} models remain below 0.2~eV, which is lower than those of the other single-task models. In particular, {\gnnnano} reproduces the Henry coefficient ($K_{\mathrm{H}}$) of CO$_2$, calculated by 10,000 insertion iterations following the ODAC25 protocol~\cite{b_mof_odac25_sriram2025open}, in better agreement with experimental data. Meanwhile, the MAE for the heat capacity is slightly higher than that of the other MLIPs, although it still remains within an acceptable range.

We investigate molecular adsorption on metallic surfaces, which is of interest in chemical processes such as catalysis and molecular conversion~\cite{msur_norskov2009towards}. As shown in \autoref{fig:bench_nano}e, the {\gnnnano} models exhibit significantly improved performance compared to {\gnnzero} and MACE-mp-0-small on the ADS41 dataset~\cite{b_ads41_mallikarjun2019adsorption,b_ads41_trepte2022data}, which includes 15 physisorption and 26 chemisorption. We explicitly exclude Co and Ni surfaces from the chemisorption evaluations due to the errors from the mpa channel of {\gnnomni}~\cite{omni_kim2025optimizing}. We further evaluate the adsorption energies of H, O, OH, and CO on five noble metals—Cu, Pd, Pt, Ag, and Au—and observe similar trends (\autoref{fig:bench_nano}e, bottom-left panel). In addition, we examine reaction energies for the adsorption of Si-centered inhibitor molecules, (\textit{N},\textit{N}-dimethylamino)trimethylsilane (DMATMS) and ethyltrichlorosilane (ETS) on SiO$_2$ and Si$_3$N$_4$, which is relevant to semiconductor processes such as area-selective deposition (ASD)~\cite{asd_lee2025molecular}. The results show that {\gnnnano} exhibits performance comparable to that of {\gnnzero} and MACE-mp-0-small in this case.

Overall, {\gnnnano}, trained on datasets generated by {\gnnomni} across a wide range of domains, benefits from the rich representations of diverse chemical systems provided by the teacher model. As a result, the {\gnnnano} models achieve broader domain coverage than {\gnnzero} and MACE-mp-0-small. This is further supported by benchmarks on the domain-bridging set (DBS), which was constructed to address the lack of DFT data for configurations evaluated with a specific exchange–correlation functional in a previous study~\cite{omni_kim2025optimizing}. For example, the DBS includes PBE results for configurations from the OC20 dataset, which were originally computed using RPBE. As shown in Figure~S2, the {\gnnnano} models generally exhibit improved performance on the PBE-calculated DBS, particularly for systems involving molecules and surfaces. Regarding the impact of $r_c$, we find no strong dependence of the results on the cutoff radius. This suggests that simply increasing $r_c$, without enhancing architectural complexity, does not significantly improve the performance of lightweight uMLIPs trained on large-scale datasets.

\subsection{Applications to materials research}
\subsubsection{Li diffusion in solid electrolytes}
\begin{figure}[!t]
\centering
\includegraphics[width=1.0\linewidth]{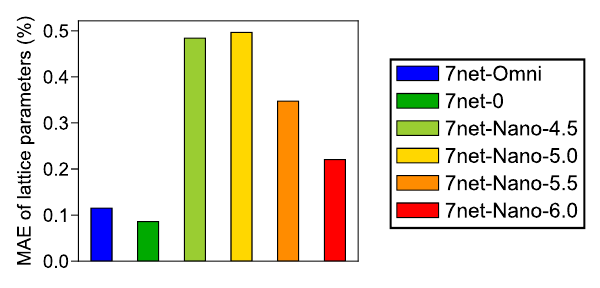}
\caption{MAEs of lattice parameters (\%) for eight argyrodite and eight non-argyrodite SSEs relaxed using {\gnnomni}, {\gnnzero}, and {\gnnnano}. The MAE is computed by evaluating the relative errors of the three lattice parameters with respect to the DFT values for each material and then averaging over all materials.} 
\label{fig:Li_lattice_parms}
\end{figure}

We demonstrate the applicability of {\gnnnano} to materials research, along with a fine-tuning approach to complement its performance when necessary. The target simulations are selected to evaluate various dynamical properties in addition to static ones. In the following, we focus on SevenNet models for benchmarking {\gnnnano}, as the MACE model is expected to exhibit performance comparable to that of {\gnnzero}, as inferred from evaluations on preexisting datasets (\autoref{fig:bench_nano}). We first examine Li transport in SSEs, which are important for developing next-generation Li-ion batteries with enhanced safety and higher energy density~\cite{sse_zhao2020designing}. \autoref{fig:Li_lattice_parms} illustrates the MAEs of the optimized lattice parameters for the 16 SSEs obtained using {\gnnomni}, {\gnnzero}, and {\gnnnano} relative to the reference DFT calculations. {\gnnomni} and {\gnnzero} yield low MAEs of $\sim$0.1{\%}. The {\gnnnano} models produce slightly larger MAEs, but they remain below 0.5{\%}, indicating excellent accuracy in predicting the lattice parameters of the tested SSEs. 

\begin{figure*}[!t]
\centering
\includegraphics[width=1.0\linewidth]{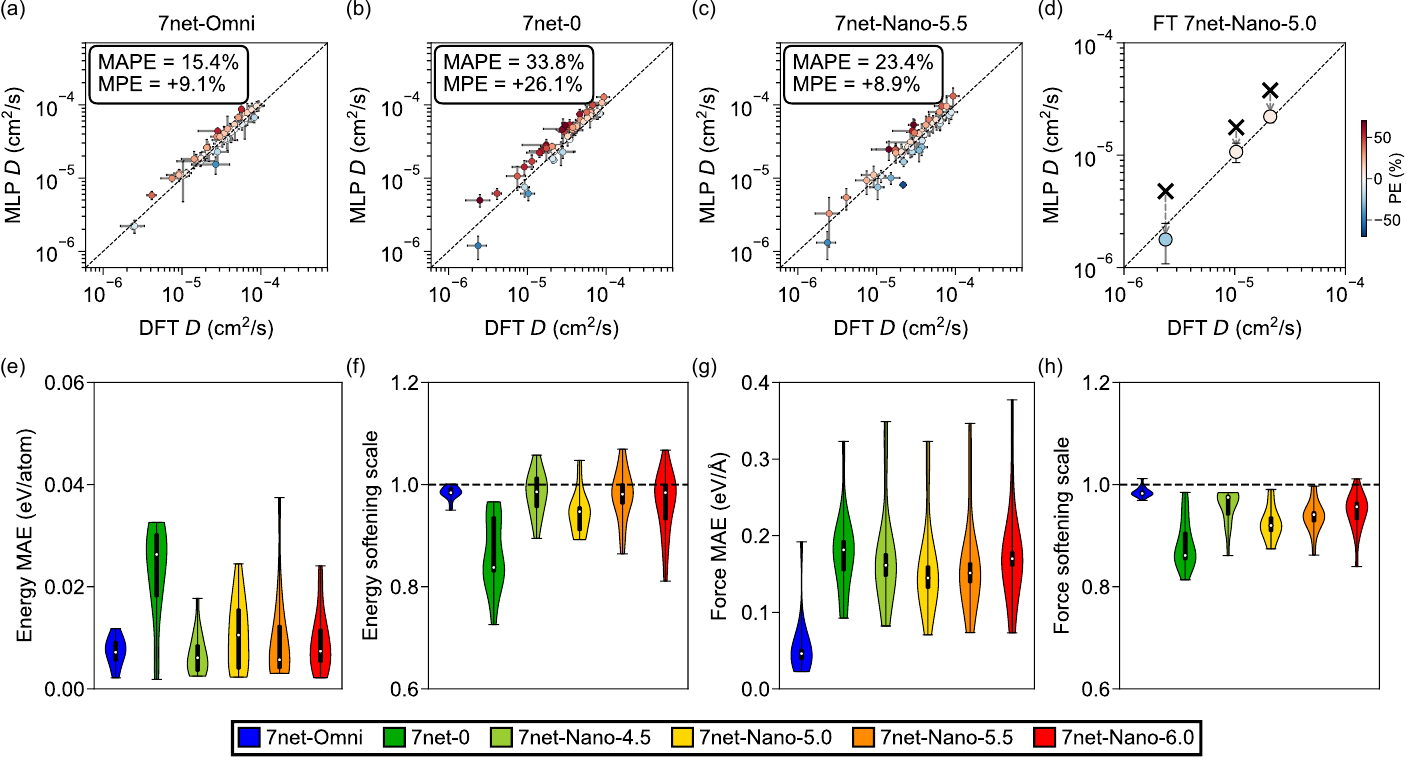}
\caption{Li-ion diffusivities of 16 target SSEs obtained from (a) {\gnnomni}, (b) {\gnnzero}, (c) {\gnnnano}-5.5 ($r_c = 5.5$~\AA), and (d) fine-tuned {\gnnnano}-5.0 ($r_c = 5.0$~\AA), compared with DFT results. Distributions of (e) energy MAEs, (f) energy softening scales, (g) force MAEs, and (h) force softening scales for the MLIPs are also shown. The DFT diffusivities and trajectories used to compute the MLIP MAEs are adopted from Ref.~\cite{ft_kim2025efficient}.}
\label{fig:Li_diffusion}
\end{figure*}

\autoref{fig:Li_diffusion}a-c show the Li-ion diffusivity in the SSEs calculated using MLIP-MD, along with DFT-MD results \cite{ft_kim2025efficient}. The mean percentage error (MPE) and mean absolute percentage error (MAPE) relative to DFT values are provided to assess both the prediction trend and accuracy. {\gnnomni} shows excellent performance in predicting Li-ion diffusivity, in good agreement with DFT results and outperforming {\gnnzero}. Its MAPE (15.4{\%}) and MPE (9.1{\%}) are considerably lower than those of {\gnnzero} (MAPE = 33.8{\%} and MPE = 26.1{\%}). In particular, {\gnnzero} exhibits a large positive MPE, indicating a systematic overestimation of Li-ion diffusivity due to the known softening behavior of universal MLIPs trained on low-energy crystal data~\cite{ftsoft_deng2025systematic}, as shown in \autoref{fig:Li_diffusion}b. In contrast, {\gnnomni} significantly mitigates this issue, yielding an MPE that is approximately three times smaller. {\gnnnano}-5.5 provides reasonable estimates to DFT results, with a slightly larger MAPE compared to that of the teacher model. It is worth noting that the MPE of {\gnnnano}-5.5 remains comparable to that of {\gnnomni}, indicating that the softening issue is effectively mitigated in the model. A similar diffusivity trend is consistently observed across {\gnnnano} models with different $r_c$ values (Figure~S3). As found in the preceding benchmarks for static properties, $r_c$ does not systematically influence the performance of {\gnnnano}. 

We further evaluate the prediction accuracy of MLIP models by examining the distribution of energy and force MAEs for the tested SSEs, as well as the resulting softening scales, defined as the slopes of linear fits between MLIP and DFT values for the energy and force of each material (\autoref{fig:Li_diffusion}e–h). The softening scale tends to decrease below unity as the PES softening becomes more pronounced. {\gnnomni} yields the lowest energy and force MAEs averaged over the tested SSEs, with softening scales close to unity on average. In contrast, {\gnnzero} exhibits the largest average energy and force MAEs and the lowest average softening scales. Compared to {\gnnzero}, the student models, {\gnnnano}, show improved accuracy, yielding lower average energy and force MAEs and softening scales closer to unity, consistent with the diffusivity analysis.

Although the present {\gnnnano} models exhibit reasonable performance in estimating Li-ion diffusivities, particularly considering their significantly lower computational cost (up to an order of magnitude, as discussed later), their accuracy can be further improved without increasing architectural complexity and computational expense through fine-tuning. For example, among the {\gnnnano} models, the {\gnnnano}-5.0 yields a large MPE of 84.6{\%} for LSN (see Table~S4 for the MAE and MAPE of each material and {\gnnmlips}). To improve its accuracy, we perform fine-tuning of the model. When constructing the fine-tuning dataset, we sample configurations from MD trajectories generated by the {\gnnnano} model itself, rather than generating them from separate teacher-model or DFT simulations~\cite{ftset_roth2026fine,ftset_kaur2025data}. This is because {\gnnnano} mostly produces physically relevant trajectories, while still including inaccurate configurations that require correction. By reusing configurations from prior simulations for sampling, we therefore reduce both computational cost and data-generation overhead. Sampling is performed from NVT MD simulations at intervals of 0.2 ps over a duration of 0.1 ns, yielding 501 configurations per trajectory and a total of 7,515 configurations from five independent simulations. 

The fine-tuning dataset is constructed by collecting energies, forces, stresses, and atomic energies from single-point {\gnnomni} calculations on the sampled MD snapshots, allocating 90\% of the dataset to the training set and 10\% to the validation set. During fine-tuning, {\gnnnano}-5.0 is retrained for 100 epochs, starting from its pretrained weights without additional constraints. We employ the Adam optimizer together with a warm-up scheduler, in which the learning rate is linearly increased to 0.03 over the first 20 epochs, followed by cosine annealing to zero over the remaining 80 epochs. This fine-tuning procedure is also applied to other applications, which will be discussed later.

\autoref{fig:Li_diffusion}d shows that the fine-tuned {\gnnnano} model (FT-{\gnnnano}-5.0) exhibits notable improvements in predicting Li-ion diffusivity, with a substantial reduction in the LSN-specific MAPE (from 84.6 to 22.4{\%}) and MPE (from 84.6 to 13.6{\%}). Notably, due to the extensive pretrained knowledge of the model, {\gnnnano} can achieve substantial accuracy improvements with a relatively small fine-tuning dataset. We indeed observe satisfactory performance gains using 10{\%} of the full fine-tuning dataset (Figure~S4). This data efficiency is particularly important for large systems, where the generation and learning of fine-tuning data can be computationally expensive and time-consuming. 

\subsubsection{Solvents for Li-ion liquid electrolytes}

\begin{figure}[!t]
\centering
\includegraphics[width=0.95\linewidth]{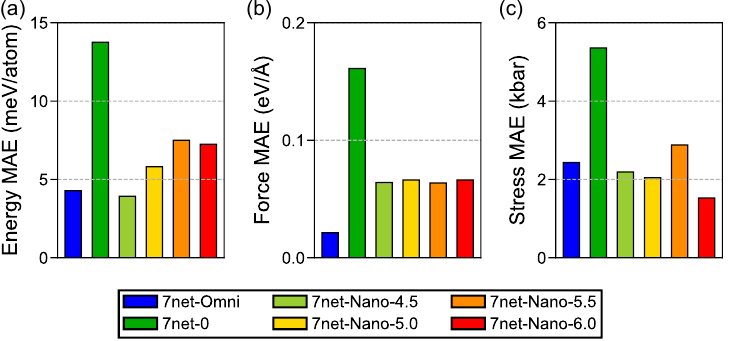}
\caption{MAEs of (a) energy, (b) force, and (c) normal stress for MLIPs, evaluated relative to DFT results for 20 Li-ion liquid electrolyte solvents. Both MLIP and DFT calculations include van der Waals interactions via the D3 method. The reference DFT+D3 data are adopted from Ref.~\cite{ft_Li_ju2025application}.}
\label{fig:liq_elect_mae}
\end{figure}

We simulate twenty solvents used as liquid electrolytes for Li-ion transport between the anode and cathode in Li-ion batteries, including cyclic and linear carbonates, esters, and ethers. The physical properties of these solvents have been extensively examined using classical and MLIP-based MD simulations in a previous study~\cite{ft_Li_ju2025application}. We first evaluate the prediction accuracy of {\gnnzero}, {\gnnomni}, and {\gnnnano} for energies, forces, and normal stress components. To this end, snapshots are sampled from 0.4-ns equilibrium simulations performed using {\gnnzero} at intervals of 4 ps. For each configuration, the corresponding quantities are computed using the MLIPs with D3 van der Waals corrections and compared with PBE+D3 results \cite{ft_Li_ju2025application}. As shown in \autoref{fig:liq_elect_mae}, {\gnnomni} exhibits excellent performance, consistently yielding low MAEs across all three metrics. The {\gnnnano} models also achieve reasonably low MAEs and, in particular, outperform {\gnnzero}, highlighting their superior representational capability for organic systems.

\begin{figure}[!t]
\centering
\includegraphics[width=1.0\linewidth]{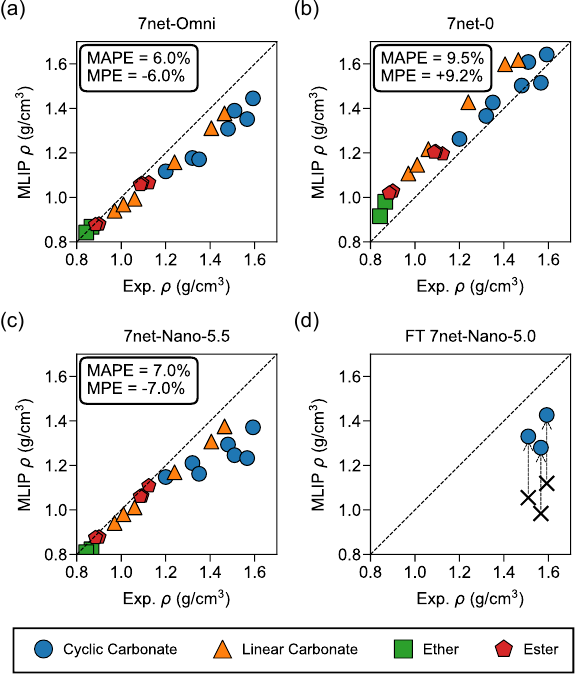}
\caption{Comparison of experimental and calculated equilibrium densities of 20 Li-ion liquid electrolyte solvents, including cyclic carbonates, linear carbonates, ethers, and esters, obtained using (a) {\gnnomni}, (b) {\gnnzero}, and (c) {\gnnnano}-5.5 ($r_c = 5.5$~\AA). (d) Equilibrium densities of DFEC, \textit{cis}-DFEC, and \textit{trans}-DFEC obtained using fine-tuned {\gnnnano}-5.0. Corresponding densities from the original {\gnnnano}-5.0 model prior to fine-tuning are denoted by $\times$. Results for {\gnnzero} are taken from Ref.~[\citenum{liqexp_xu2004nonaqueous,liqexp_gong2025predictive,liqexp_almasi2023density,liqexp_francca2009influence,liqexp_guard1999chemical,liqexp_hagiyama2008physical,liqexp_vali2016vinylene,liqexp_yang2007study,liqexp_zheng2008density}].}
\label{fig:liq_elect}
\end{figure}

\autoref{fig:liq_elect} compares the equilibrium densities of the solvents, obtained by averaging the instantaneous densities during MD simulations, using {\gnnomni}, {\gnnzero}, and {\gnnnano}-5.5 with experimental values (the results for {\gnnnano} with other $r_c$ values are provided in Figure~S5 of the Supporting Information). Across all solvents, {\gnnzero} tends to systematically overestimate the densities, yielding a MAPE of 9.5{\%}, similar to an MPE of 9.2{\%}. This behavior is attributed to its limited accuracy in describing intermolecular interactions and its tendency to overestimate stresses, compared to PBE+D3, which is expected to reasonably predict experimental densities~\cite{ft_Li_ju2025application}. In contrast, {\gnnomni} better reproduces the experimental densities, although underestimations are observed, particularly for high-density solvents of cyclic and linear carbonates. Analysis of the stress parity (normal components) between {\gnnomni} and PBE+D3 shows that {\gnnomni} tends to predict larger stress values, which favor larger equilibrium volumes (i.e., lower densities). Despite the underestimation of densities, we emphasize that the predictive quality of MD simulations using {\gnnomni} is excellent, as evidenced by the low MAPE of 6{\%} and its consistency with experimental density trends across the solvents. {\gnnnano} reproduces density trends comparable to those of {\gnnomni} with smaller stress MAE than {\gnnzero} (\autoref{fig:liq_elect_mae}), as shown in \autoref{fig:liq_elect}c, although some deviations are observed depending on the $r_c$ value (Figure~S5). The stress parities of {\gnnnano} also closely follow those of its teacher model, as shown in Figure~S7--10.

As in the case of Li transport, fine-tuning can be efficiently applied to improve the accuracy of {\gnnnano} for solvent systems. For example, among the student models, {\gnnnano}-5.0 exhibits relatively large discrepancies in the predicted densities of carbonate solvents, such as DFEC, \textit{cis}-DFEC, and \textit{trans}-DFEC, compared to those obtained with {\gnnomni} (see Table~S5 for the solvent densities predicted by different MLIP methods). For its fine-tuning, configurations are sampled from corresponding {\gnnnano} MD simulations at intervals of 1 ps during the 0.4-ns equilibration run for the target systems. A total of 401 configurations are collected for each of DFEC, \textit{cis}-DFEC, and \textit{trans}-DFEC, yielding 1,203 structures in total. The fine-tuning dataset is then constructed by collecting {\gnnomni} results for these configurations. Fine-tuning proceeds following the same procedure used for the Li solid-state electrolytes described above. As shown in \autoref{fig:liq_elect}d, fine-tuning effectively improves agreement with the density values obtained from {\gnnomni} simulations. The MAEs in energies and forces are also reduced to levels comparable to those of {\gnnomni} (Figure~S11). In addition, the fine-tuned model reproduces stress distributions during MD simulations consistent with those of {\gnnomni} (Figure~S12).

\subsubsection{Plasma etching of SiO$_2$}
We simulate plasma etching of SiO$_2$, which is a critical process in semiconductor device fabrication~\cite{har_cao2023future,sio2etch_hayashi1996char,sio2etch_lin2018achieving,sio2etch_lee2010ultrahigh}, using MLIPs. From a modeling perspective, this simulation is significantly more challenging than property predictions, such as those for Li electrolytes discussed above, because it involves a wide range of chemical interactions at the surface induced by the incidence of high-energy species (up to a few hundred eV). Therefore, this simulation requires substantially higher accuracy from MLIPs in terms of chemical-space coverage—encompassing molecules, bulks, surfaces, and molecule–surface interactions. Furthermore, during high-energy impacts, interatomic distances can shrink well below 1~\AA. Thus, short-range repulsive interactions, which are often poorly described by MLIPs trained primarily on low-energy crystal data, must be accurately captured. Indeed, etching simulations are not successfully performed using {\gnnzero} due to its inability to describe such short-range interactions, which will be discussed in detail below. Besides accuracy considerations, obtaining physically meaningful results requires large system sizes and long simulation times, which can be computationally demanding even for MLIPs if their computational cost is not sufficiently low. 

\begin{figure}[!t]
\includegraphics[width=\columnwidth]{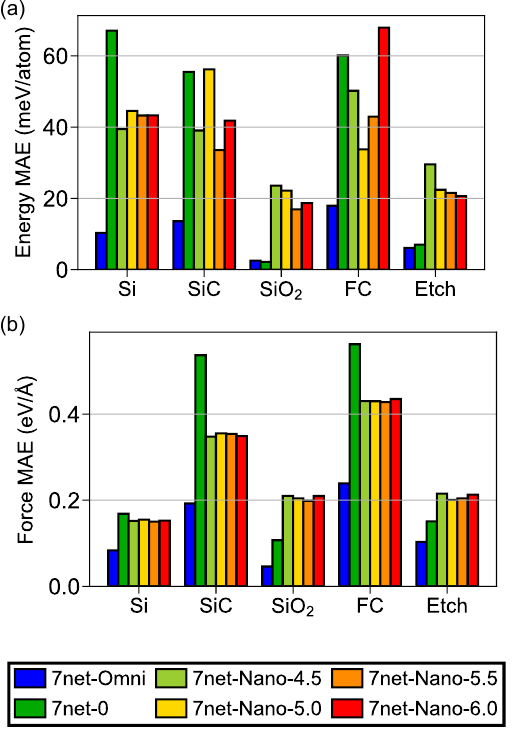}
\caption{MAEs of (a) energy and (b) force for MLIPs evaluated on structures sampled from melt–quench–anneal simulations of \ch{Si}, \ch{SiC}, \ch{SiO2}, and fluorocarbons (FCs), as well as from etching trajectories generated using {\gnnomni}. All errors are evaluated relative to the corresponding DFT results.} 
\label{fig:etch_oneshot_DFT}
\end{figure}

We first evaluate the accuracy of {\gnnomni}, {\gnnzero}, and {\gnnnano} for amorphous systems associated with \ch{SiO2} etching, such as Si, SiC, \ch{SiO2}, and fluorocarbons (FCs), as well as for etching trajectories, by comparing their predictions with reference DFT calculations. Configurations for this evaluation are sampled from MD simulations performed using {\gnnomni}. For bulk systems, configurations are extracted from melt–quench–anneal trajectories. Etching simulations for data sampling in this benchmarking are performed using a small surface model with an area of 1~nm$^2$ to ensure the feasibility of DFT calculations. The kinetic energy of incident \ch{CF2} and \ch{CF3} molecules is limited to 30~eV due to the small surface area; higher incident energies would lead to excessive energy transfer per unit area upon impact, causing total melting of the system. We also assess the ability of the MLIPs to describe short-range interactions by examining the PES as a function of interatomic distance. The corresponding configurations are obtained from quasi-static drag (QSD) calculations, in which a CF molecule is gradually driven toward a target atom—either Si or O on the \ch{SiO2} surface, or residual C or F atoms present during etching. Detailed sampling procedures for this benchmarking are provided in Supplementary Note~1.

In \autoref{fig:etch_oneshot_DFT}, {\gnnomni} exhibits excellent accuracy across all tested systems and etching configurations, with energy and force MAEs below 20 meV/atom and 0.25 eV/\AA, respectively. In contrast, {\gnnzero} generally shows larger MAEs for bulk systems, except for \ch{SiO2}, where the energy and force MAEs remain close to those of {\gnnomni}. We find that {\gnnzero} also yields relatively small MAEs for etching trajectories; however, this does not necessarily indicate an accurate description of the etching process. This agreement is limited to the mild impact conditions employed in the present benchmarking, where the kinetic energies of incident species are restricted to 30~eV. Consequently, interatomic distances upon impact do not become sufficiently short to expose the intrinsic errors of {\gnnzero}. Its accuracy deteriorates markedly under more severe impact conditions relevant to actual plasma etching, where the kinetic energies reach hundreds of electron volts. For the {\gnnnano} models, the energy and force MAEs are smaller than or comparable to those of {\gnnzero} under these benchmarking conditions.

\begin{figure}[!t]
\centering
\includegraphics[width=\columnwidth]{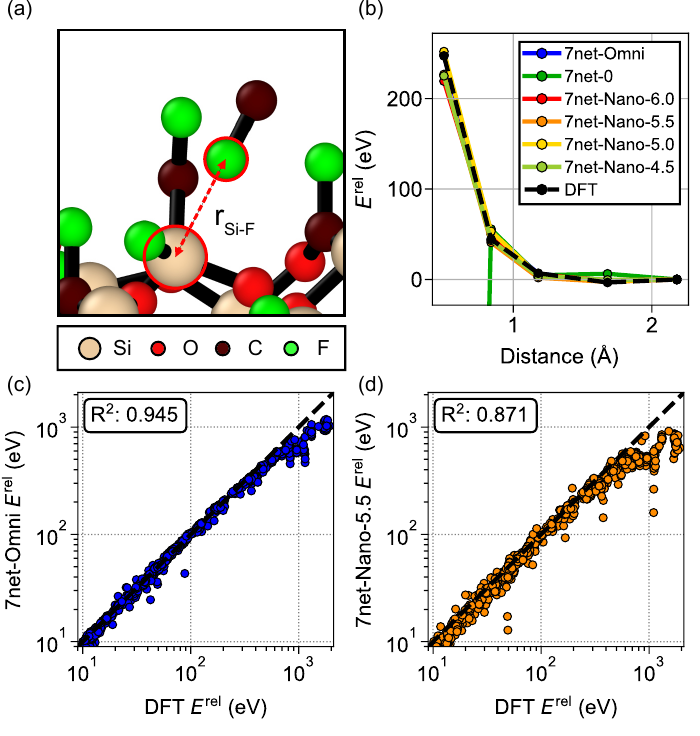}
\caption{Quasi-static drag (QSD) calculations between the surface and a \ch{CF} molecule. (a) Representative structure in which the \ch{F} atom of \ch{CF} is dragged toward a \ch{Si} atom on the substrate. (b) DFT and MLIP predictions for energy variation as a function of the interatomic distance for the configuration in (a). In (b), $E^{\mathrm{rel}}$ is defined by setting the energy of the configuration at the largest interatomic distance (corresponding to $1.3 \times$ the sum of covalent radii) to zero. Scatter plots of $E^{\mathrm{rel}}$ from the QSD calculations are shown for (c) {\gnnomni} and (d) {\gnnnano}-5.5, based on 1,075 configurations with $E^{\mathrm{rel}} > 10$~eV.} 
\label{fig:etch_qsd}
\end{figure}

As mentioned above, stable etching simulations require MLIPs to reliably capture short-range interactions, which we assess using QSD simulations. We consider 268 QSD simulations with different orientations of an incoming CF molecule, along with surface configurations extracted from the etching trajectories of {\gnnomni}. A representative structure and the corresponding energy profiles as a function of the interatomic distance between the F atom in the CF molecule and a surface Si atom are shown in \autoref{fig:etch_qsd}a and b, respectively. In \autoref{fig:etch_qsd}b, we present the relative energies ($E^{\mathrm{rel}}$), referenced to the configuration at the largest interatomic distance, corresponding to 1.3 times the sum of covalent radii, used in the QSD simulations. In the DFT calculations, it is evident that repulsive interactions become dominant as the interatomic distance decreases, leading to an increase in energy up to several hundred electron volts. Notably, both {\gnnomni} and {\gnnnano} accurately capture this trend. In contrast, {\gnnzero} exhibits unphysical behavior, with the energy abruptly decreasing to negative values below a critical distance ($\sim$0.8~\AA), rendering it unsuitable for plasma etching simulations involving high-energy ion impacts.

To further confirm the robustness of {\gnnomni} and {\gnnnano} for short-range interactions, we select configurations from 268 QSD simulations that yield $E^{\mathrm{rel}}$ values between 10 and 2000 eV in DFT calculations and compute the corresponding $E^{\mathrm{rel}}$ values using the MLIP models. \autoref{fig:etch_qsd}c and d show scatter plots of $E^{\mathrm{rel}}$ for {\gnnomni} and {\gnnnano}-5.5, respectively, relative to the DFT references (results for the other {\gnnnano} models are provided in Figure~S13). {\gnnomni} accurately reproduces the DFT repulsive interaction energies, yielding an excellent $R^2$ value of 0.945, with only slight underestimation for $E^{\mathrm{rel}}$ above 1000 eV. This demonstrates the capability of {\gnnomni} to simulate etching events involving high-energy collisions without requiring additional short-range corrections. {\gnnnano} also shows reasonable accuracy, with an $R^2$ value above 0.871. Taken together with the low energy and force MAEs across diverse systems and configurations (\autoref{fig:etch_oneshot_DFT}), which are comparable to those of MLIPs previously used to simulate etching processes~\cite{chfetch_An_2026,hfetch_Hong_2024}, this strong capability to capture short-range interactions suggests that, like {\gnnomni}, the student models enable stable MD simulations under severe etching conditions.

We note that the high accuracy of {\gnnomni} in describing short-range interactions arises from its enhanced representation across a wide range of interatomic distances, enabled by the inclusion of high-energy training data from OMat24 containing configurations with very short interatomic separations. During distillation, this capability is effectively transferred to the student model {\gnnnano}. In contrast, the MPtrj training dataset used for {\gnnzero} lacks such high-energy configurations, limiting its ability to represent short-range interactions. 

We also separately evaluate the accuracy of the MLIPs for diverse molecular species associated with \ch{SiO2} etching, ranging from stable byproducts to reactive radicals, as listed in Table~S6. Both {\gnnomni} and its student models show excellent agreement with DFT: the energy MAE is 0.06 eV/atom for {\gnnomni} and 0.16–0.20 eV/atom for the student models. The $R^2$ values are also close to unity (above 0.95; Figure~S14), indicating the high accuracy of the MLIPs.

\begin{figure}[t]
\centering
\includegraphics[width=1.0\linewidth]{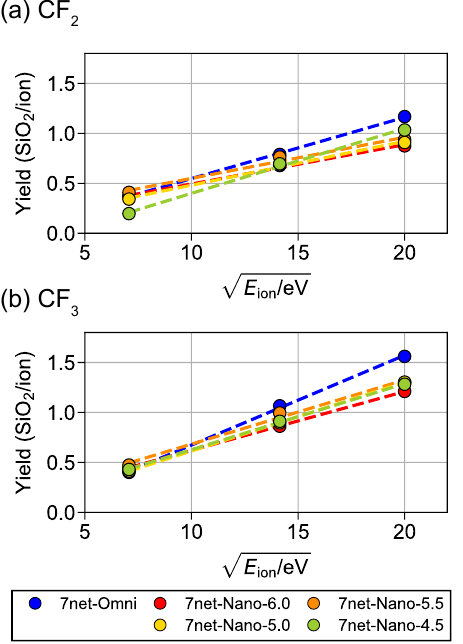}
\caption{\ch{SiO2} etching yields from simulations using (a) \ch{CF2} and (b) \ch{CF3} at 50, 200, and 400~eV, obtained with {\gnnomni} and {\gnnnano}.} 
\label{fig:etch_yield_pretrain}
\end{figure}

\autoref{fig:etch_yield_pretrain} illustrates the \ch{SiO2} etching yield from MLIP-MD simulations using a substrate with an area of 4~nm$^2$. During etching, the actual ratio of removed Si to O atoms is not exactly 1:2, but remains close to this stoichiometry. Therefore, we estimate the \ch{SiO2} etching yield by approximating the number of removed formula units as the total number of removed atoms divided by three. Among the total 800 impact events, we consider only the last 400 events (corresponding to an ion dose of 10$^{16}$~cm$^{-2}$), after the system reaches a steady state. For both \ch{CF2} and \ch{CF3} impacts, the \ch{SiO2} etching yield is found to be proportional to the square root of the incident ion energy, consistent with experimental observations~\cite{etch_rel_steinbruchel1989universal,ionbeam_karahashi2004etching}. The {\gnnnano} models reproduce trends similar to those of {\gnnomni}, with a slight underestimation at high impact energies, demonstrating their strong capability as pretrained student models. 

Fine-tuning can be applied to further enhance the accuracy of the {\gnnnano} models. We fine-tune the {\gnnnano}-5.0 using configurations obtained from its etching trajectories. Specifically, configurations are sampled every 400 steps during the first 600 \ch{CF2} and \ch{CF3} ion impacts (out of a total of 800) for each incident energy. This sampling procedure yields 30,188 snapshots, with the number of atoms per structure ranging from approximately 1,800 to 2,000. We then perform single-point {\gnnomni} calculations for each sampled configuration to generate the fine-tuning dataset. Because using the full dataset is computationally demanding and increases training time, we screen the configurations to select those with large errors for inclusion in the fine-tuning dataset. It should be noted that, because collision events in etching processes are highly localized at the surface, global error metrics such as the MAE of forces or per-atom energies are not suitable for sampling in this case. In addition, the total number of atoms varies within approximately 10{\%} during etching, making it difficult to define consistent error thresholds based on normalized quantities. Accordingly, we focus on individual atomic forces and select configurations in which the maximum absolute force error exceeds \SI{2}{eV/\angstrom}, considering the error distribution (Figure~S15). As an additional criterion to account for energy errors, we impose a tolerance of 2~eV on atomic energy differences. This strategy leverages a key advantage of the student models, which directly learn atomic energies from the teacher model. To further reduce the dataset size, we restrict sampling to configurations extracted from simulations with incident energies of 200 and 400 eV, which effectively capture the large-error configurations observed in the 50 eV case. The final fine-tuning dataset comprises 17,392 configurations.

During fine-tuning, we additionally employ a replay approach, in which a portion of the pretrained dataset is used together with the target-specific fine-tuning dataset. This approach mitigates catastrophic forgetting, which would otherwise significantly degrade simulation performance~\cite{ft_kim2025efficient}. Unlike the previous examples involving Li diffusion and solvent properties, catastrophic forgetting must be treated with particular care in etching simulations, where it can degrade the accuracy for complex surface structures and molecular interactions during ion impacts, particularly short-range repulsions, which are otherwise well captured by the student models. We observe that a fine-tuned {\gnnnano} model without replay produces unphysical configurations and degrades accuracy for short-range interactions, a behavior not observed prior to fine-tuning (see Supplementary Note~2 for details). For the replay application, the replay set is constructed by randomly selecting 1 million configurations from OMat24 and 100,000 configurations from each of the MPtrj, Alex, OC22, QCML, and SPICE datasets. The fine-tuning process is then carried out by subsampling the replay set to maintain a 1:20 ratio between the fine-tuning dataset and the subsampled replay set, accounting for differences in the average number of atoms per structure between etching trajectories and configurations in the replay dataset.

\begin{figure}[!t]
\centering
\includegraphics[width=1.0\linewidth]{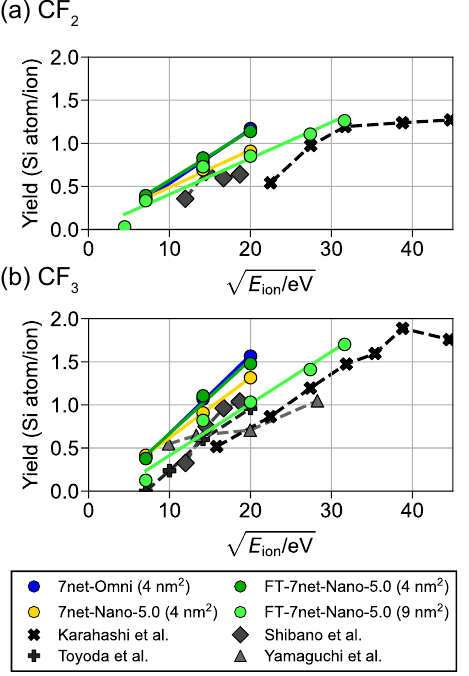}
\caption{Si yield from \ch{SiO2} etching simulations using (a) \ch{CF2} and (b) \ch{CF3}, comparing {\gnnomni}, {\gnnnano}-5.0, and fine-tuned {\gnnnano}-5.0 (FT-7net-Nano-5.0). All three MLIPs simulate etching using a 4~nm$^2$ surface model, while the fine-tuned {\gnnnano} additionally employs a 9~nm$^2$ surface model. Experimental ion-beam data from Refs.~\cite{ionbeam_toyoda2004beam,ionbeam_shibano1993etching,ionbeam_yamaguchi2000etching,ionbeam_karahashi2004etching} are included for comparison.}
\label{fig:yield_3nm}
\end{figure}

With the replay method, the fine-tuned {\gnnnano} model (FT-{\gnnnano}-5.0) successfully simulates etching processes without exhibiting unphysical structures. In \autoref{fig:yield_3nm}, the etching yield of Si atoms is compared across simulations and ion-beam experiments~\cite{ionbeam_toyoda2004beam,ionbeam_shibano1993etching,ionbeam_yamaguchi2000etching,ionbeam_karahashi2004etching}. Notably, FT-{\gnnnano}-5.0 accurately reproduces the results obtained from {\gnnomni} using a 4-nm$^2$ substrate model. Compared to experiments, simulations using the 4-nm$^2$ model overestimate the Si yield. To examine the cell-size effect, we employ a larger substrate with an area of 9~nm$^2$, which is expected to yield a lower etching yield by reducing the impact energy density and facilitating energy dissipation due to the increased number of atoms in the system. In this case, simulations are performed using FT-{\gnnnano}-5.0, while {\gnnomni} is not applied due to its high computational cost. The larger simulation scale enables the use of higher impact energies (750 and 1000~eV) for both \ch{CF2} and \ch{CF3}. These conditions cannot be properly simulated using the 4-nm$^2$ model, as the substrate undergoes melting due to the high impact energy density. We extract the etching yield from the last 900 impacts out of 1800 in total, after confirming that the system has reached a steady state. As expected, the 9-nm$^2$ model leads to a lower etching yield at a given impact energy, improving agreement with experimental observations. We also confirm that the simulated surface composition aligns excellently with experimental data, as shown in Figure~S16. This result clearly demonstrates the effectiveness of {\gnnnano}, achieving both computational efficiency and high accuracy.

\subsection{Speedup tests of {\gnnnano}} 

To evaluate the computational throughput of {\gnnnano} relative to other MLIP models, we perform NVT MD simulations for amorphous \ch{SiO2} systems of varying sizes at 300~K. We first generate a 70-atom amorphous \ch{SiO2} structure with a 1-nm$^3$ cubic cell, which is subsequently replicated to construct models containing between 70 and 70,000 atoms. The simulations are carried out for 2~ps with a time step of 1~fs on an NVIDIA RTX PRO 6000 GPU (80~GB). For comparison, we evaluate the throughput of {\gnnomni}, {\gnnzero}, and MACE-mp-0-small~\cite{mace_batatia2025foundation} with cuEquivariance (v0.9.1) \cite{cueq_geiger2024} under the same simulation conditions.

\begin{figure}[!t]
\centering
\includegraphics[width=1.0\linewidth]{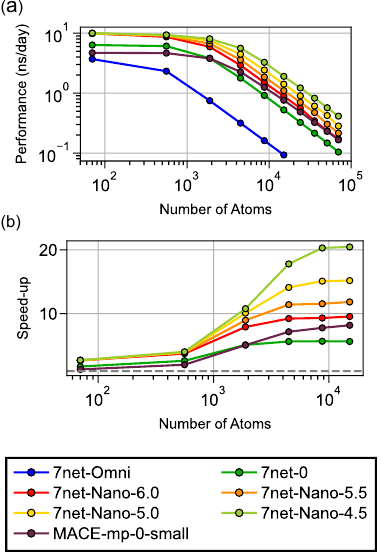}
\caption{Speedup tests for amorphous \ch{SiO2} MD simulations with system sizes ranging from 70 to 70,000 atoms on an NVIDIA RTX PRO 6000 (80~GB). The performance of {\gnnomni}, all {\gnnnano} variants, and MACE-mp-0-small is compared in terms of (a) nanoseconds per day and (b) speedup relative to {\gnnomni}. Data points for {\gnnomni} beyond 15,120 atoms are omitted in (a) due to out-of-memory issues.}
\label{fig:time_perf}
\end{figure}

As shown in \autoref{fig:time_perf}a, the throughput of {\gnnomni} decreases from \SI{3.695}{ns/day} for 70 atoms to \SI{0.093}{ns/day} for 15,120 atoms, beyond which out-of-memory errors occur. Similarly, the throughput of {\gnnzero} decreases from \SI{6.356}{ns/day} for 70 atoms to \SI{0.525}{ns/day} for 15,120 atoms, and further to \SI{0.104}{ns/day} for 70,000 atoms. MACE-mp-0-small exhibits a reduction in throughput from 4.703 to 0.166 ns/day as the system size increases from 70 to 70,000 atoms, showing performance comparable to or slightly better than that of {\gnnzero} at large system sizes. In contrast, the {\gnnnano}-6.0 maintains the higher throughput, ranging from \SI{9.895}{ns/day} to \SI{0.170}{ns/day} for systems containing 70 to 70,000 atoms, respectively. The other {\gnnnano} models with smaller $r_c$ values achieve even higher throughputs. 

The variation in computational cost among the {\gnnnano} models arises primarily from neighbor list construction, which depends on the cutoff distance and increasingly impacts performance as system size grows to some extent. Specifically, for a system with 70,000 atoms, the {\gnnnano} models with $r_c$ values of 5.5, 5.0, and 4.5\AA\ are more productive than the $r_c=6.0$ \AA\ model by factors of 1.26, 1.66, and 2.43, respectively.

From the throughput test, we quantify the computational speedup of {\gnnnano} relative to its teacher model (\autoref{fig:time_perf}b). For small systems (fewer than 1,000 atoms), the speedup relative to {\gnnomni} is approximately 2.68, largely independent of $r_c$. However, as the system size increases, the speedup becomes more pronounced. For systems exceeding 10,000 atoms, the speedup relative to {\gnnomni} reaches 9.5, 11.8, 15.2, and 20.45 for $r_c$ values of 6.0, 5.5, 5.0, and 4.5 \AA\, respectively. These results, together with much smaller memory overhead, clearly demonstrate that {\gnnnano} accelerates simulations by an order of magnitude or more, making it highly suitable for large-scale applications.

\subsection{Implications and Outlook for {\gnnnano}}
As demonstrated in the preceding discussions, {\gnnnano} is a lightweight uMLIP developed via knowledge distillation from {\gnnomni}, enabling large-scale simulations with reduced computational cost and high simulation stability. To further illustrate the computational efficiency of {\gnnnano}, we compare its simulation time with that of SIMPLE-NN, a descriptor-based bespoke MLIP known for its superior efficiency relative to GNN-based uMLIPs. For \ch{SiO2} etching simulations at a similar scale (9~nm$^2$ surface area), the {\gnnnano}-5.5 model achieves a simulation speed of 26.8 steps/sec on an NVIDIA RTX PRO 6000 GPU, which is within the same order of magnitude as the 86.8 steps/sec obtained from SIMPLE-NN simulations that we performed on two nodes with 24 CPU cores (Intel(R) Xeon(R) Gold 6226), following the procedure described in a previous study~\cite{chfetch_An_2026}. This result demonstrates that {\gnnnano} is an efficient GNN-based pretrained MLIP. Moreover, the computational cost of GNN-based models is largely insensitive to the number of elements. As such, they can offer a clear advantage over descriptor-based approaches for large-scale simulations involving multiple elements, where descriptor-based MLIPs incur increased computational cost.

Recently, LightPFP, another pretrained student model derived from the PFP teacher model, has been reported~\cite{lightpfp_li2025}. As it is based on MTP descriptors, it is expected to achieve faster simulations than {\gnnnano}. However, in contrast to LightPFP—where fine-tuning to specific applications, often iteratively depending on system complexity, is typically required—GNN-based {\gnnnano} enables stable MD simulations across a wide range of applications without fine-tuning and remains robust as the simulation scale and chemical complexity increase. When necessary, fine-tuning can also be applied to {\gnnnano} with relatively low additional effort.

Finally, as shown above, the accuracy of {\gnnnano} exhibits weak dependence on $r_c$. Among the developed models, {\gnnnano}-5.5 provides consistently reliable performance across the benchmarks and serves as a robust pretrained model. Therefore, it would be a reasonable choice. For applications requiring higher computational efficiency, the lighter {\gnnnano}-4.5 model can be employed, with optional fine-tuning if needed.

\section{Conclusions}
In this work, we develop a lightweight universal machine-learning interatomic potential, {\gnnnano}, based on the SevenNet architecture and enabled by knowledge distillation from the foundation model {\gnnomni}. Despite its substantially reduced model size, {\gnnnano} achieves reliable accuracy across diverse chemical domains. In particular, it provides reasonable predictions for Li-ion diffusion in solid-state electrolytes while mitigating the force-softening issue often observed in pretrained MLIPs. It also accurately reproduces the densities of liquid electrolytes obtained from the teacher model, demonstrating consistent performance for molecular systems and improving upon conventional pretrained models trained on low-energy crystal data. In addition, {\gnnnano} captures short-range repulsive interactions with high fidelity, enabling physically meaningful simulations of high-energy processes such as plasma etching of \ch{SiO2}. Importantly, {\gnnnano} delivers significant computational advantages, achieving order-of-magnitude speedup relative to its teacher model while avoiding memory limitations. This enables large-scale simulations involving complex chemistries and extended system sizes that are otherwise difficult to access with the foundation model. Furthermore, its accuracy can be enhanced, when necessary, through efficient fine-tuning with relatively small datasets. Overall, {\gnnnano} provides a practical and transferable framework for atomistic simulations, offering a balanced combination of accuracy, efficiency, and scalability. These features make the {\gnnnano} models a compelling and practical choice for large-scale simulations across a wide range of materials applications.

\section{Data Availability}
The checkpoints of {\gnnnano}, packages of SevenNet which are modified for fine-tuning,  example code to fine-tune liquid electrolyte application, and the DFT single-points calculations of the benchmark for the \ch{SiO2} with \ch{CF$_x$} plasma etching simulations, including melt-quench-annealing, fluorocarbon, plasma etching, and quasi-static drags are available on Zenodo at \url{https://zenodo.org/records/19491140}. 

\begin{suppinfo}

The following files are available free of charge.
\begin{itemize}
  \item Supporting Information: This file includes simulation details for benchmark configuration sampling in \ch{SiO2} plasma etching; Failure in plasma etching simulations; validation MAEs of {\gnnnano}, MAEs on the domain bridge set; Li-ion diffusivity of SSEs using {\gnnnano}; influence of dataset size for fine-tuning LSN; equilibrium densities of solvent using {\gnnnano}; normal stress parity of {\gnnomni}, {\gnnnano} relative to DFT; MAEs of fine-tuned {\gnnnano} relative to DFEC, \textit{cis}-DFEC, \textit{trans}-DFEC;  quasi-static drag calculations of {\gnnnano}; molecular energies of {\gnnnano}; error distribution of fine-tuning set for plasma etching; surface composition of plasma etching using fine-tuned {\gnnnano}; argyrodite compositions; simulation conditions of plasma etching; description and source of benchmark tasks; MPEs and MAPEs of each SSEs; density of each solvents; list of molecules for plasma etching benchmark
  
\end{itemize}

\end{suppinfo}

\begin{acknowledgement}
This work was supported by the National Research Foundation of Korea (NRF) grant funded by the Korea government (MSIT) (No. RS-2023-00247245 and No. RS-2024-00407840). The computations were carried out at Korea Institute of Science and Technology Information (KISTI) National Supercomputing Center (KSC-2025-CRE-0284) and at the Center for Advanced Computations (CAC) at Korea Institute for Advanced Study (KIAS).
\end{acknowledgement}

\bibliography{ref}

\providecommand{\latin}[1]{#1}
\makeatletter
\providecommand{\doi}
  {\begingroup\let\do\@makeother\dospecials
  \catcode`\{=1 \catcode`\}=2 \doi@aux}
\providecommand{\doi@aux}[1]{\endgroup\texttt{#1}}
\makeatother
\providecommand*\mcitethebibliography{\thebibliography}
\csname @ifundefined\endcsname{endmcitethebibliography}  {\let\endmcitethebibliography\endthebibliography}{}
\begin{mcitethebibliography}{118}
\providecommand*\natexlab[1]{#1}
\providecommand*\mciteSetBstSublistMode[1]{}
\providecommand*\mciteSetBstMaxWidthForm[2]{}
\providecommand*\mciteBstWouldAddEndPuncttrue
  {\def\EndOfBibitem{\unskip.}}
\providecommand*\mciteBstWouldAddEndPunctfalse
  {\let\EndOfBibitem\relax}
\providecommand*\mciteSetBstMidEndSepPunct[3]{}
\providecommand*\mciteSetBstSublistLabelBeginEnd[3]{}
\providecommand*\EndOfBibitem{}
\mciteSetBstSublistMode{f}
\mciteSetBstMaxWidthForm{subitem}{(\alph{mcitesubitemcount})}
\mciteSetBstSublistLabelBeginEnd
  {\mcitemaxwidthsubitemform\space}
  {\relax}
  {\relax}

\bibitem[Behler(2016)]{mlip_behler2016perspective}
Behler,~J. Perspective: Machine learning potentials for atomistic simulations. \emph{The Journal of Chemical Physics} \textbf{2016}, \emph{145}\relax
\mciteBstWouldAddEndPuncttrue
\mciteSetBstMidEndSepPunct{\mcitedefaultmidpunct}
{\mcitedefaultendpunct}{\mcitedefaultseppunct}\relax
\EndOfBibitem
\bibitem[Deringer \latin{et~al.}(2019)Deringer, Caro, and Cs{\'a}nyi]{mlip_deringer2019machine}
Deringer,~V.~L.; Caro,~M.~A.; Cs{\'a}nyi,~G. Machine learning interatomic potentials as emerging tools for materials science. \emph{Advanced Materials} \textbf{2019}, \emph{31}, 1902765\relax
\mciteBstWouldAddEndPuncttrue
\mciteSetBstMidEndSepPunct{\mcitedefaultmidpunct}
{\mcitedefaultendpunct}{\mcitedefaultseppunct}\relax
\EndOfBibitem
\bibitem[Behler and Parrinello(2007)Behler, and Parrinello]{bpnn_behler2007generalized}
Behler,~J.; Parrinello,~M. Generalized neural-network representation of high-dimensional potential-energy surfaces. \emph{Physical review letters} \textbf{2007}, \emph{98}, 146401\relax
\mciteBstWouldAddEndPuncttrue
\mciteSetBstMidEndSepPunct{\mcitedefaultmidpunct}
{\mcitedefaultendpunct}{\mcitedefaultseppunct}\relax
\EndOfBibitem
\bibitem[Shapeev(2016)]{mtp_shapeev2016moment}
Shapeev,~A.~V. Moment tensor potentials: A class of systematically improvable interatomic potentials. \emph{Multiscale Modeling \& Simulation} \textbf{2016}, \emph{14}, 1153--1173\relax
\mciteBstWouldAddEndPuncttrue
\mciteSetBstMidEndSepPunct{\mcitedefaultmidpunct}
{\mcitedefaultendpunct}{\mcitedefaultseppunct}\relax
\EndOfBibitem
\bibitem[Novikov \latin{et~al.}(2021)Novikov, Gubaev, Podryabinkin, and Shapeev]{mtp_novikov2021mlip}
Novikov,~I.~S.; Gubaev,~K.; Podryabinkin,~E.~V.; Shapeev,~A.~V. The MLIP package: moment tensor potentials with MPI and active learning. \emph{Machine Learning: Science and Technology} \textbf{2021}, \emph{2}, 025002\relax
\mciteBstWouldAddEndPuncttrue
\mciteSetBstMidEndSepPunct{\mcitedefaultmidpunct}
{\mcitedefaultendpunct}{\mcitedefaultseppunct}\relax
\EndOfBibitem
\bibitem[De \latin{et~al.}(2016)De, Bart{\'o}k, Cs{\'a}nyi, and Ceriotti]{soap_de2016comparing}
De,~S.; Bart{\'o}k,~A.~P.; Cs{\'a}nyi,~G.; Ceriotti,~M. Comparing molecules and solids across structural and alchemical space. \emph{Physical Chemistry Chemical Physics} \textbf{2016}, \emph{18}, 13754--13769\relax
\mciteBstWouldAddEndPuncttrue
\mciteSetBstMidEndSepPunct{\mcitedefaultmidpunct}
{\mcitedefaultendpunct}{\mcitedefaultseppunct}\relax
\EndOfBibitem
\bibitem[Rhodes \latin{et~al.}(2025)Rhodes, Vandenhaute, Šimkus, Gin, Godwin, Duignan, and Neumann]{orb_rhodes2025orb}
Rhodes,~B.; Vandenhaute,~S.; Šimkus,~V.; Gin,~J.; Godwin,~J.; Duignan,~T.; Neumann,~M. Orb-v3: atomistic simulation at scale. 2025; \url{https://arxiv.org/abs/2504.06231}\relax
\mciteBstWouldAddEndPuncttrue
\mciteSetBstMidEndSepPunct{\mcitedefaultmidpunct}
{\mcitedefaultendpunct}{\mcitedefaultseppunct}\relax
\EndOfBibitem
\bibitem[Wood \latin{et~al.}(2026)Wood, Dzamba, Fu, Gao, Shuaibi, Barroso-Luque, Abdelmaqsoud, Gharakhanyan, Kitchin, Levine, Michel, Sriram, Cohen, Das, Rizvi, Sahoo, Ulissi, and Zitnick]{uma_wood2025family}
Wood,~B.~M. \latin{et~al.}  UMA: A Family of Universal Models for Atoms. 2026; \url{https://arxiv.org/abs/2506.23971}\relax
\mciteBstWouldAddEndPuncttrue
\mciteSetBstMidEndSepPunct{\mcitedefaultmidpunct}
{\mcitedefaultendpunct}{\mcitedefaultseppunct}\relax
\EndOfBibitem
\bibitem[Zhang \latin{et~al.}(2026)Zhang, Peng, Cai, Li, Zhou, Zeng, Guo, Zhang, Li, Jiang, Zhu, Jia, Zhang, and Wang]{dpa_zhang2025graph}
Zhang,~D.; Peng,~A.; Cai,~C.; Li,~W.; Zhou,~Y.; Zeng,~J.; Guo,~M.; Zhang,~C.; Li,~B.; Jiang,~H.; Zhu,~T.; Jia,~W.; Zhang,~L.; Wang,~H. A Graph Neural Network for the Era of Large Atomistic Models. 2026; \url{https://arxiv.org/abs/2506.01686}\relax
\mciteBstWouldAddEndPuncttrue
\mciteSetBstMidEndSepPunct{\mcitedefaultmidpunct}
{\mcitedefaultendpunct}{\mcitedefaultseppunct}\relax
\EndOfBibitem
\bibitem[Fu \latin{et~al.}(2025)Fu, Wood, Barroso-Luque, Levine, Gao, Dzamba, and Zitnick]{esen_fu2025learning}
Fu,~X.; Wood,~B.~M.; Barroso-Luque,~L.; Levine,~D.~S.; Gao,~M.; Dzamba,~M.; Zitnick,~C.~L. Learning Smooth and Expressive Interatomic Potentials for Physical Property Prediction. 2025; \url{https://arxiv.org/abs/2502.12147}\relax
\mciteBstWouldAddEndPuncttrue
\mciteSetBstMidEndSepPunct{\mcitedefaultmidpunct}
{\mcitedefaultendpunct}{\mcitedefaultseppunct}\relax
\EndOfBibitem
\bibitem[Lysogorskiy \latin{et~al.}(2026)Lysogorskiy, Bochkarev, and Drautz]{b_gb_lysogorskiy2026graph}
Lysogorskiy,~Y.; Bochkarev,~A.; Drautz,~R. Graph atomic cluster expansion for foundational machine learning interatomic potentials. \emph{npj Computational Materials} \textbf{2026}, \emph{12}, 114\relax
\mciteBstWouldAddEndPuncttrue
\mciteSetBstMidEndSepPunct{\mcitedefaultmidpunct}
{\mcitedefaultendpunct}{\mcitedefaultseppunct}\relax
\EndOfBibitem
\bibitem[Yang \latin{et~al.}(2024)Yang, Hu, Zhou, Liu, Shi, Li, Li, Chen, Chen, Zeni, Horton, Pinsler, Fowler, Zügner, Xie, Smith, Sun, Wang, Kong, Liu, Hao, and Lu]{mattersim_yang2024mattersim}
Yang,~H. \latin{et~al.}  MatterSim: A Deep Learning Atomistic Model Across Elements, Temperatures and Pressures. 2024; \url{https://arxiv.org/abs/2405.04967}\relax
\mciteBstWouldAddEndPuncttrue
\mciteSetBstMidEndSepPunct{\mcitedefaultmidpunct}
{\mcitedefaultendpunct}{\mcitedefaultseppunct}\relax
\EndOfBibitem
\bibitem[Batatia \latin{et~al.}(2025)Batatia, Lin, Hart, Kasoar, Elena, Norwood, Wolf, and Csányi]{macemh_batatia2025cross}
Batatia,~I.; Lin,~C.; Hart,~J.; Kasoar,~E.; Elena,~A.~M.; Norwood,~S.~W.; Wolf,~T.; Csányi,~G. Cross Learning between Electronic Structure Theories for Unifying Molecular, Surface, and Inorganic Crystal Foundation Force Fields. 2025; \url{https://arxiv.org/abs/2510.25380}\relax
\mciteBstWouldAddEndPuncttrue
\mciteSetBstMidEndSepPunct{\mcitedefaultmidpunct}
{\mcitedefaultendpunct}{\mcitedefaultseppunct}\relax
\EndOfBibitem
\bibitem[Kim \latin{et~al.}(2026)Kim, You, Park, Lim, Kang, Kim, Jeon, Ju, Hong, Lee, Choi, Kim, Lee, and Han]{omni_kim2025optimizing}
Kim,~J.; You,~J.; Park,~Y.; Lim,~Y.; Kang,~Y.; Kim,~J.; Jeon,~H.; Ju,~S.; Hong,~D.; Lee,~S.~Y.; Choi,~S.; Kim,~Y.; Lee,~J.~W.; Han,~S. Optimizing cross-domain transfer for universal machine learning interatomic potentials. 2026; \url{https://doi.org/10.1038/s41467-026-70195-8}\relax
\mciteBstWouldAddEndPuncttrue
\mciteSetBstMidEndSepPunct{\mcitedefaultmidpunct}
{\mcitedefaultendpunct}{\mcitedefaultseppunct}\relax
\EndOfBibitem
\bibitem[Qu and Krishnapriyan(2024)Qu, and Krishnapriyan]{scale_qu2024importance}
Qu,~E.; Krishnapriyan,~A.~S. The importance of being scalable: Improving the speed and accuracy of neural network interatomic potentials across chemical domains. \emph{Advances in Neural Information Processing Systems} \textbf{2024}, \emph{37}, 139030--139053\relax
\mciteBstWouldAddEndPuncttrue
\mciteSetBstMidEndSepPunct{\mcitedefaultmidpunct}
{\mcitedefaultendpunct}{\mcitedefaultseppunct}\relax
\EndOfBibitem
\bibitem[Liao \latin{et~al.}(2024)Liao, Wood, Das, and Smidt]{scale_liao2023equiformerv2}
Liao,~Y.-L.; Wood,~B.; Das,~A.; Smidt,~T. EquiformerV2: Improved Equivariant Transformer for Scaling to Higher-Degree Representations. 2024; \url{https://arxiv.org/abs/2306.12059}\relax
\mciteBstWouldAddEndPuncttrue
\mciteSetBstMidEndSepPunct{\mcitedefaultmidpunct}
{\mcitedefaultendpunct}{\mcitedefaultseppunct}\relax
\EndOfBibitem
\bibitem[Leimeroth \latin{et~al.}(2025)Leimeroth, Erhard, Albe, and Rohrer]{scale_leimeroth2025machine}
Leimeroth,~N.; Erhard,~L.~C.; Albe,~K.; Rohrer,~J. Machine-learning interatomic potentials from a user's perspective: a comparison of accuracy, speed and data efficiency. \emph{Modelling and Simulation in Materials Science and Engineering} \textbf{2025}, \emph{33}, 065012\relax
\mciteBstWouldAddEndPuncttrue
\mciteSetBstMidEndSepPunct{\mcitedefaultmidpunct}
{\mcitedefaultendpunct}{\mcitedefaultseppunct}\relax
\EndOfBibitem
\bibitem[Xie \latin{et~al.}(2023)Xie, Rupp, and Hennig]{fast_xie2023ultra}
Xie,~S.~R.; Rupp,~M.; Hennig,~R.~G. Ultra-fast interpretable machine-learning potentials. \emph{npj Computational Materials} \textbf{2023}, \emph{9}, 162\relax
\mciteBstWouldAddEndPuncttrue
\mciteSetBstMidEndSepPunct{\mcitedefaultmidpunct}
{\mcitedefaultendpunct}{\mcitedefaultseppunct}\relax
\EndOfBibitem
\bibitem[Kong \latin{et~al.}(2026)Kong, Shim, Hu, and Fung]{fast_kong2026scalable}
Kong,~L.; Shim,~J.; Hu,~G.; Fung,~V. Scalable foundation interatomic potentials via message-passing pruning and graph partitioning. 2026; \url{https://doi.org/10.1038/s41524-026-02001-4}\relax
\mciteBstWouldAddEndPuncttrue
\mciteSetBstMidEndSepPunct{\mcitedefaultmidpunct}
{\mcitedefaultendpunct}{\mcitedefaultseppunct}\relax
\EndOfBibitem
\bibitem[Taniguchi(2025)]{mixedloss_taniguchi2025knowledge}
Taniguchi,~T. Knowledge distillation of neural network potential for molecular crystals. \emph{Faraday Discussions} \textbf{2025}, \emph{256}, 139--155\relax
\mciteBstWouldAddEndPuncttrue
\mciteSetBstMidEndSepPunct{\mcitedefaultmidpunct}
{\mcitedefaultendpunct}{\mcitedefaultseppunct}\relax
\EndOfBibitem
\bibitem[Jung(2025)]{ae_jung2025atomic}
Jung,~G.~S. Atomic Energy Accuracy of Neural Network Potentials: Harnessing Pretraining and Transfer Learning. \emph{Journal of Chemical Information and Modeling} \textbf{2025}, \emph{65}, 4797--4807\relax
\mciteBstWouldAddEndPuncttrue
\mciteSetBstMidEndSepPunct{\mcitedefaultmidpunct}
{\mcitedefaultendpunct}{\mcitedefaultseppunct}\relax
\EndOfBibitem
\bibitem[Matin \latin{et~al.}(2025)Matin, Allen, Shinkle, Pachalieva, Craven, Nebgen, Smith, Messerly, Li, Tretiak, Barros, and Lubbers]{ae_Matin_2025}
Matin,~S.; Allen,~A. E.~A.; Shinkle,~E.; Pachalieva,~A.; Craven,~G.~T.; Nebgen,~B.; Smith,~J.~S.; Messerly,~R.; Li,~Y.~W.; Tretiak,~S.; Barros,~K.; Lubbers,~N. Teacher-student training improves the accuracy and efficiency of machine learning interatomic potentials. \emph{Digital Discovery} \textbf{2025}, \emph{4}, 2502–2511\relax
\mciteBstWouldAddEndPuncttrue
\mciteSetBstMidEndSepPunct{\mcitedefaultmidpunct}
{\mcitedefaultendpunct}{\mcitedefaultseppunct}\relax
\EndOfBibitem
\bibitem[Amin \latin{et~al.}(2025)Amin, Raja, and Krishnapriyan]{hessian_Sanjeev_2025}
Amin,~I.; Raja,~S.; Krishnapriyan,~A. Towards Fast, Specialized Machine Learning Force Fields: Distilling Foundation Models via Energy Hessians. 2025; \url{https://arxiv.org/abs/2501.09009}\relax
\mciteBstWouldAddEndPuncttrue
\mciteSetBstMidEndSepPunct{\mcitedefaultmidpunct}
{\mcitedefaultendpunct}{\mcitedefaultseppunct}\relax
\EndOfBibitem
\bibitem[Ekstr{\"o}m~Kelvinius \latin{et~al.}(2023)Ekstr{\"o}m~Kelvinius, Georgiev, Toshev, and Gasteiger]{feature_ekstrom2023accelerating}
Ekstr{\"o}m~Kelvinius,~F.; Georgiev,~D.; Toshev,~A.; Gasteiger,~J. Accelerating molecular graph neural networks via knowledge distillation. \emph{Advances in Neural Information Processing Systems} \textbf{2023}, \emph{36}, 25761--25792\relax
\mciteBstWouldAddEndPuncttrue
\mciteSetBstMidEndSepPunct{\mcitedefaultmidpunct}
{\mcitedefaultendpunct}{\mcitedefaultseppunct}\relax
\EndOfBibitem
\bibitem[Li \latin{et~al.}(2025)Li, Charoenphakdee, Zhuang, Okuno, Tsuboi, Takamoto, Ishida, and Li]{lightpfp_li2025}
Li,~W.; Charoenphakdee,~N.; Zhuang,~Y.-B.; Okuno,~R.; Tsuboi,~Y.; Takamoto,~S.; Ishida,~J.; Li,~J. LightPFP: A Lightweight Route to Ab Initio Accuracy at Scale. 2025; \url{https://arxiv.org/abs/2510.23064}\relax
\mciteBstWouldAddEndPuncttrue
\mciteSetBstMidEndSepPunct{\mcitedefaultmidpunct}
{\mcitedefaultendpunct}{\mcitedefaultseppunct}\relax
\EndOfBibitem
\bibitem[Takamoto \latin{et~al.}(2022)Takamoto, Shinagawa, Motoki, Nakago, Li, Kurata, Watanabe, Yayama, Iriguchi, Asano, and {et al.}]{pfp_takamoto2022towards}
Takamoto,~S.; Shinagawa,~C.; Motoki,~D.; Nakago,~K.; Li,~W.; Kurata,~I.; Watanabe,~T.; Yayama,~Y.; Iriguchi,~H.; Asano,~Y.; {et al.} Towards universal neural network potential for material discovery applicable to arbitrary combination of 45 elements. \emph{Nature Communications} \textbf{2022}, \emph{13}, 2991\relax
\mciteBstWouldAddEndPuncttrue
\mciteSetBstMidEndSepPunct{\mcitedefaultmidpunct}
{\mcitedefaultendpunct}{\mcitedefaultseppunct}\relax
\EndOfBibitem
\bibitem[Loshchilov and Hutter(2019)Loshchilov, and Hutter]{adamw_loshchilov2017decoupled}
Loshchilov,~I.; Hutter,~F. Decoupled Weight Decay Regularization. 2019; \url{https://arxiv.org/abs/1711.05101}\relax
\mciteBstWouldAddEndPuncttrue
\mciteSetBstMidEndSepPunct{\mcitedefaultmidpunct}
{\mcitedefaultendpunct}{\mcitedefaultseppunct}\relax
\EndOfBibitem
\bibitem[Deng \latin{et~al.}(2023)Deng, Zhong, Jun, Riebesell, Han, Bartel, and Ceder]{mptrj_deng2023chgnet}
Deng,~B.; Zhong,~P.; Jun,~K.; Riebesell,~J.; Han,~K.; Bartel,~C.~J.; Ceder,~G. CHGNet as a pretrained universal neural network potential for charge-informed atomistic modelling. \emph{Nature Machine Intelligence} \textbf{2023}, \emph{5}, 1031--1041\relax
\mciteBstWouldAddEndPuncttrue
\mciteSetBstMidEndSepPunct{\mcitedefaultmidpunct}
{\mcitedefaultendpunct}{\mcitedefaultseppunct}\relax
\EndOfBibitem
\bibitem[Kingsbury \latin{et~al.}(2022)Kingsbury, Gupta, Bartel, Munro, Dwaraknath, Horton, and Persson]{mpr2scan_kingsbury2022performance}
Kingsbury,~R.; Gupta,~A.~S.; Bartel,~C.~J.; Munro,~J.~M.; Dwaraknath,~S.; Horton,~M.; Persson,~K.~A. Performance comparison of r 2 SCAN and SCAN metaGGA density functionals for solid materials via an automated, high-throughput computational workflow. \emph{Physical Review Materials} \textbf{2022}, \emph{6}, 013801\relax
\mciteBstWouldAddEndPuncttrue
\mciteSetBstMidEndSepPunct{\mcitedefaultmidpunct}
{\mcitedefaultendpunct}{\mcitedefaultseppunct}\relax
\EndOfBibitem
\bibitem[Merchant \latin{et~al.}(2023)Merchant, Batzner, Schoenholz, Aykol, Cheon, and Cubuk]{mpr2scan_merchant2023scaling}
Merchant,~A.; Batzner,~S.; Schoenholz,~S.~S.; Aykol,~M.; Cheon,~G.; Cubuk,~E.~D. Scaling deep learning for materials discovery. \emph{Nature} \textbf{2023}, \emph{624}, 80--85\relax
\mciteBstWouldAddEndPuncttrue
\mciteSetBstMidEndSepPunct{\mcitedefaultmidpunct}
{\mcitedefaultendpunct}{\mcitedefaultseppunct}\relax
\EndOfBibitem
\bibitem[Schmidt \latin{et~al.}(2023)Schmidt, Hoffmann, Wang, Borlido, Carri{\c{c}}o, Cerqueira, Botti, and Marques]{alex_schmidt2023machine}
Schmidt,~J.; Hoffmann,~N.; Wang,~H.-C.; Borlido,~P.; Carri{\c{c}}o,~P.~J.; Cerqueira,~T.~F.; Botti,~S.; Marques,~M.~A. Machine-learning-assisted determination of the global zero-temperature phase diagram of materials. \emph{Advanced Materials} \textbf{2023}, \emph{35}, 2210788\relax
\mciteBstWouldAddEndPuncttrue
\mciteSetBstMidEndSepPunct{\mcitedefaultmidpunct}
{\mcitedefaultendpunct}{\mcitedefaultseppunct}\relax
\EndOfBibitem
\bibitem[Schmidt \latin{et~al.}(2024)Schmidt, Cerqueira, Romero, Loew, J{\"a}ger, Wang, Botti, and Marques]{alex_schmidt2024improving}
Schmidt,~J.; Cerqueira,~T.~F.; Romero,~A.~H.; Loew,~A.; J{\"a}ger,~F.; Wang,~H.-C.; Botti,~S.; Marques,~M.~A. Improving machine-learning models in materials science through large datasets. \emph{Materials Today Physics} \textbf{2024}, \emph{48}, 101560\relax
\mciteBstWouldAddEndPuncttrue
\mciteSetBstMidEndSepPunct{\mcitedefaultmidpunct}
{\mcitedefaultendpunct}{\mcitedefaultseppunct}\relax
\EndOfBibitem
\bibitem[Barroso-Luque \latin{et~al.}(2024)Barroso-Luque, Shuaibi, Fu, Wood, Dzamba, Gao, Rizvi, Zitnick, and Ulissi]{omat24_barroso2024open}
Barroso-Luque,~L.; Shuaibi,~M.; Fu,~X.; Wood,~B.~M.; Dzamba,~M.; Gao,~M.; Rizvi,~A.; Zitnick,~C.~L.; Ulissi,~Z.~W. Open Materials 2024 (OMat24) Inorganic Materials Dataset and Models. 2024; \url{https://arxiv.org/abs/2410.12771}\relax
\mciteBstWouldAddEndPuncttrue
\mciteSetBstMidEndSepPunct{\mcitedefaultmidpunct}
{\mcitedefaultendpunct}{\mcitedefaultseppunct}\relax
\EndOfBibitem
\bibitem[Kaplan \latin{et~al.}(2025)Kaplan, Liu, Qi, Ko, Deng, Riebesell, Ceder, Persson, and Ong]{matpes_kaplan2025foundational}
Kaplan,~A.~D.; Liu,~R.; Qi,~J.; Ko,~T.~W.; Deng,~B.; Riebesell,~J.; Ceder,~G.; Persson,~K.~A.; Ong,~S.~P. A Foundational Potential Energy Surface Dataset for Materials. 2025; \url{https://arxiv.org/abs/2503.04070}\relax
\mciteBstWouldAddEndPuncttrue
\mciteSetBstMidEndSepPunct{\mcitedefaultmidpunct}
{\mcitedefaultendpunct}{\mcitedefaultseppunct}\relax
\EndOfBibitem
\bibitem[Levine \latin{et~al.}(2026)Levine, Shuaibi, Spotte-Smith, Taylor, Hasyim, Michel, Batatia, Csányi, Dzamba, Eastman, Frey, Fu, Gharakhanyan, Krishnapriyan, Rackers, Raja, Rizvi, Rosen, Ulissi, Vargas, Zitnick, Blau, and Wood]{omol25_levine2025open}
Levine,~D.~S. \latin{et~al.}  The Open Molecules 2025 (OMol25) Dataset, Evaluations, and Models. 2026; \url{https://arxiv.org/abs/2505.08762}\relax
\mciteBstWouldAddEndPuncttrue
\mciteSetBstMidEndSepPunct{\mcitedefaultmidpunct}
{\mcitedefaultendpunct}{\mcitedefaultseppunct}\relax
\EndOfBibitem
\bibitem[Eastman \latin{et~al.}(2023)Eastman, Behara, Dotson, Galvelis, Herr, Horton, Mao, Chodera, Pritchard, Wang, and {et al.}]{spice_eastman2023}
Eastman,~P.; Behara,~P.~K.; Dotson,~D.~L.; Galvelis,~R.; Herr,~J.~E.; Horton,~J.~T.; Mao,~Y.; Chodera,~J.~D.; Pritchard,~B.~P.; Wang,~Y.; {et al.} Spice, a dataset of drug-like molecules and peptides for training machine learning potentials. \emph{Scientific Data} \textbf{2023}, \emph{10}, 11\relax
\mciteBstWouldAddEndPuncttrue
\mciteSetBstMidEndSepPunct{\mcitedefaultmidpunct}
{\mcitedefaultendpunct}{\mcitedefaultseppunct}\relax
\EndOfBibitem
\bibitem[Ganscha \latin{et~al.}(2025)Ganscha, Unke, Ahlin, Maennel, Kashubin, and M{\"u}ller]{qcml_ganscha2025}
Ganscha,~S.; Unke,~O.~T.; Ahlin,~D.; Maennel,~H.; Kashubin,~S.; M{\"u}ller,~K.-R. The QCML dataset, Quantum chemistry reference data from 33.5 M DFT and 14.7 B semi-empirical calculations. \emph{Scientific Data} \textbf{2025}, \emph{12}, 406\relax
\mciteBstWouldAddEndPuncttrue
\mciteSetBstMidEndSepPunct{\mcitedefaultmidpunct}
{\mcitedefaultendpunct}{\mcitedefaultseppunct}\relax
\EndOfBibitem
\bibitem[Chanussot \latin{et~al.}(2021)Chanussot, Das, Goyal, Lavril, Shuaibi, Riviere, Tran, Heras-Domingo, Ho, Hu, and {et al.}]{oc20_chanussot2021open}
Chanussot,~L.; Das,~A.; Goyal,~S.; Lavril,~T.; Shuaibi,~M.; Riviere,~M.; Tran,~K.; Heras-Domingo,~J.; Ho,~C.; Hu,~W.; {et al.} Open catalyst 2020 (OC20) dataset and community challenges. \emph{Acs Catalysis} \textbf{2021}, \emph{11}, 6059--6072\relax
\mciteBstWouldAddEndPuncttrue
\mciteSetBstMidEndSepPunct{\mcitedefaultmidpunct}
{\mcitedefaultendpunct}{\mcitedefaultseppunct}\relax
\EndOfBibitem
\bibitem[Tran \latin{et~al.}(2023)Tran, Lan, Shuaibi, Wood, Goyal, Das, Heras-Domingo, Kolluru, Rizvi, Shoghi, and {et al.}]{oc22_tran2023open}
Tran,~R.; Lan,~J.; Shuaibi,~M.; Wood,~B.~M.; Goyal,~S.; Das,~A.; Heras-Domingo,~J.; Kolluru,~A.; Rizvi,~A.; Shoghi,~N.; {et al.} The Open Catalyst 2022 (OC22) dataset and challenges for oxide electrocatalysts. \emph{ACS Catalysis} \textbf{2023}, \emph{13}, 3066--3084\relax
\mciteBstWouldAddEndPuncttrue
\mciteSetBstMidEndSepPunct{\mcitedefaultmidpunct}
{\mcitedefaultendpunct}{\mcitedefaultseppunct}\relax
\EndOfBibitem
\bibitem[Sriram \latin{et~al.}(2024)Sriram, Choi, Yu, Brabson, Das, Ulissi, Uyttendaele, Medford, and Sholl]{odac23_sriram2024open}
Sriram,~A.; Choi,~S.; Yu,~X.; Brabson,~L.~M.; Das,~A.; Ulissi,~Z.; Uyttendaele,~M.; Medford,~A.~J.; Sholl,~D.~S. The Open DAC 2023 dataset and challenges for sorbent discovery in direct air capture. 2024\relax
\mciteBstWouldAddEndPuncttrue
\mciteSetBstMidEndSepPunct{\mcitedefaultmidpunct}
{\mcitedefaultendpunct}{\mcitedefaultseppunct}\relax
\EndOfBibitem
\bibitem[Mazitov \latin{et~al.}(2025)Mazitov, Chorna, Fraux, Bercx, Pizzi, De, and Ceriotti]{mad_mazitov2025massive}
Mazitov,~A.; Chorna,~S.; Fraux,~G.; Bercx,~M.; Pizzi,~G.; De,~S.; Ceriotti,~M. Massive Atomic Diversity: a compact universal dataset for atomistic machine learning. \emph{Scientific data} \textbf{2025}, \emph{12}, 1857\relax
\mciteBstWouldAddEndPuncttrue
\mciteSetBstMidEndSepPunct{\mcitedefaultmidpunct}
{\mcitedefaultendpunct}{\mcitedefaultseppunct}\relax
\EndOfBibitem
\bibitem[Cohen \latin{et~al.}(2025)Cohen, Riebesell, Goodall, Kolluru, Falletta, Krause, Colindres, Ceder, and Gangan]{torchsim_cohen2025}
Cohen,~O.; Riebesell,~J.; Goodall,~R.; Kolluru,~A.; Falletta,~S.; Krause,~J.; Colindres,~J.; Ceder,~G.; Gangan,~A.~S. TorchSim: An efficient atomistic simulation engine in PyTorch. \emph{AI for Science} \textbf{2025}, \emph{1}, 025003\relax
\mciteBstWouldAddEndPuncttrue
\mciteSetBstMidEndSepPunct{\mcitedefaultmidpunct}
{\mcitedefaultendpunct}{\mcitedefaultseppunct}\relax
\EndOfBibitem
\bibitem[Thompson \latin{et~al.}(2022)Thompson, Aktulga, Berger, Bolintineanu, Brown, Crozier, In't~Veld, Kohlmeyer, Moore, Nguyen, and {et al.}]{lammps_thompson2022}
Thompson,~A.~P.; Aktulga,~H.~M.; Berger,~R.; Bolintineanu,~D.~S.; Brown,~W.~M.; Crozier,~P.~S.; In't~Veld,~P.~J.; Kohlmeyer,~A.; Moore,~S.~G.; Nguyen,~T.~D.; {et al.} LAMMPS-a flexible simulation tool for particle-based materials modeling at the atomic, meso, and continuum scales. \emph{Computer physics communications} \textbf{2022}, \emph{271}, 108171\relax
\mciteBstWouldAddEndPuncttrue
\mciteSetBstMidEndSepPunct{\mcitedefaultmidpunct}
{\mcitedefaultendpunct}{\mcitedefaultseppunct}\relax
\EndOfBibitem
\bibitem[Lee \latin{et~al.}(2025)Lee, Kim, Park, Jeong, Han, Park, and Lee]{leeflashtp}
Lee,~S.~Y.; Kim,~H.; Park,~Y.; Jeong,~D.; Han,~S.; Park,~Y.; Lee,~J.~W. Flashtp: Fused, sparsity-aware tensor product for machine learning interatomic potentials. Forty-second International Conference on Machine Learning. 2025\relax
\mciteBstWouldAddEndPuncttrue
\mciteSetBstMidEndSepPunct{\mcitedefaultmidpunct}
{\mcitedefaultendpunct}{\mcitedefaultseppunct}\relax
\EndOfBibitem
\bibitem[He \latin{et~al.}(2018)He, Zhu, Epstein, and Mo]{sse_ox_he2018statistical}
He,~X.; Zhu,~Y.; Epstein,~A.; Mo,~Y. Statistical variances of diffusional properties from ab initio molecular dynamics simulations. \emph{npj Computational Materials} \textbf{2018}, \emph{4}, 18\relax
\mciteBstWouldAddEndPuncttrue
\mciteSetBstMidEndSepPunct{\mcitedefaultmidpunct}
{\mcitedefaultendpunct}{\mcitedefaultseppunct}\relax
\EndOfBibitem
\bibitem[Murugan \latin{et~al.}(2007)Murugan, Thangadurai, and Weppner]{sse_ox_murugan2007fast}
Murugan,~R.; Thangadurai,~V.; Weppner,~W. Fast lithium ion conduction in garnet-type Li7La3Zr2O12. \emph{Angewandte Chemie International Edition} \textbf{2007}, \emph{46}, 7778--7781\relax
\mciteBstWouldAddEndPuncttrue
\mciteSetBstMidEndSepPunct{\mcitedefaultmidpunct}
{\mcitedefaultendpunct}{\mcitedefaultseppunct}\relax
\EndOfBibitem
\bibitem[Kamaya \latin{et~al.}(2011)Kamaya, Homma, Yamakawa, Hirayama, Kanno, Yonemura, Kamiyama, Kato, Hama, Kawamoto, and {et al.}]{sse_sul_kamaya2011lithium}
Kamaya,~N.; Homma,~K.; Yamakawa,~Y.; Hirayama,~M.; Kanno,~R.; Yonemura,~M.; Kamiyama,~T.; Kato,~Y.; Hama,~S.; Kawamoto,~K.; {et al.} A lithium superionic conductor. \emph{Nature Materials} \textbf{2011}, \emph{10}, 682--686\relax
\mciteBstWouldAddEndPuncttrue
\mciteSetBstMidEndSepPunct{\mcitedefaultmidpunct}
{\mcitedefaultendpunct}{\mcitedefaultseppunct}\relax
\EndOfBibitem
\bibitem[Seino \latin{et~al.}(2014)Seino, Ota, Takada, Hayashi, and Tatsumisago]{sse_sul_seino2014sulphide}
Seino,~Y.; Ota,~T.; Takada,~K.; Hayashi,~A.; Tatsumisago,~M. A sulphide lithium super ion conductor is superior to liquid ion conductors for use in rechargeable batteries. \emph{Energy \& Environmental Science} \textbf{2014}, \emph{7}, 627--631\relax
\mciteBstWouldAddEndPuncttrue
\mciteSetBstMidEndSepPunct{\mcitedefaultmidpunct}
{\mcitedefaultendpunct}{\mcitedefaultseppunct}\relax
\EndOfBibitem
\bibitem[Asano \latin{et~al.}(2018)Asano, Sakai, Ouchi, Sakaida, Miyazaki, and Hasegawa]{sse_hal_asano2018solid}
Asano,~T.; Sakai,~A.; Ouchi,~S.; Sakaida,~M.; Miyazaki,~A.; Hasegawa,~S. Solid halide electrolytes with high lithium-ion conductivity for application in 4 V class bulk-type all-solid-state batteries. \emph{Advanced Materials} \textbf{2018}, \emph{30}, 1803075\relax
\mciteBstWouldAddEndPuncttrue
\mciteSetBstMidEndSepPunct{\mcitedefaultmidpunct}
{\mcitedefaultendpunct}{\mcitedefaultseppunct}\relax
\EndOfBibitem
\bibitem[Li \latin{et~al.}(2010)Li, Zhu, and Chen]{sse_nit_li2010li+}
Li,~W.; Zhu,~Y.; Chen,~G. Li+ ion conductivity and diffusion mechanism in $\alpha$-Li3N and $\beta$-Li3N. \emph{Energy \& Environmental Science} \textbf{2010}, \emph{3}, 1524--1530\relax
\mciteBstWouldAddEndPuncttrue
\mciteSetBstMidEndSepPunct{\mcitedefaultmidpunct}
{\mcitedefaultendpunct}{\mcitedefaultseppunct}\relax
\EndOfBibitem
\bibitem[Miara \latin{et~al.}(2015)Miara, Suzuki, Richards, Wang, Kim, and Ceder]{sse_nit_miara2015li}
Miara,~L.~J.; Suzuki,~N.; Richards,~W.~D.; Wang,~Y.; Kim,~J.~C.; Ceder,~G. Li-ion conductivity in Li9S3N. \emph{Journal of Materials Chemistry A} \textbf{2015}, \emph{3}, 20338--20344\relax
\mciteBstWouldAddEndPuncttrue
\mciteSetBstMidEndSepPunct{\mcitedefaultmidpunct}
{\mcitedefaultendpunct}{\mcitedefaultseppunct}\relax
\EndOfBibitem
\bibitem[Kim \latin{et~al.}(2025)Kim, Lee, Oh, Park, Hwang, Han, Kang, and Kang]{ft_kim2025efficient}
Kim,~J.; Lee,~J.; Oh,~S.; Park,~Y.; Hwang,~S.; Han,~S.; Kang,~S.; Kang,~Y. An efficient forgetting-aware fine-tuning framework for pretrained universal machine-learning interatomic potentials. 2025\relax
\mciteBstWouldAddEndPuncttrue
\mciteSetBstMidEndSepPunct{\mcitedefaultmidpunct}
{\mcitedefaultendpunct}{\mcitedefaultseppunct}\relax
\EndOfBibitem
\bibitem[Hjorth~Larsen \latin{et~al.}(2017)Hjorth~Larsen, J{\o}rgen~Mortensen, Blomqvist, Castelli, Christensen, Du{\l}ak, Friis, Groves, Hammer, Hargus, and {et al.}]{ase_hjorth2017atomic}
Hjorth~Larsen,~A.; J{\o}rgen~Mortensen,~J.; Blomqvist,~J.; Castelli,~I.~E.; Christensen,~R.; Du{\l}ak,~M.; Friis,~J.; Groves,~M.~N.; Hammer,~B.; Hargus,~C.; {et al.} The atomic simulation environment—a Python library for working with atoms. \emph{Journal of Physics: Condensed Matter} \textbf{2017}, \emph{29}, 273002\relax
\mciteBstWouldAddEndPuncttrue
\mciteSetBstMidEndSepPunct{\mcitedefaultmidpunct}
{\mcitedefaultendpunct}{\mcitedefaultseppunct}\relax
\EndOfBibitem
\bibitem[Ju \latin{et~al.}(2025)Ju, You, Kim, Park, An, and Han]{ft_Li_ju2025application}
Ju,~S.; You,~J.; Kim,~G.; Park,~Y.; An,~H.; Han,~S. Application of pretrained universal machine-learning interatomic potential for physicochemical simulation of liquid electrolytes in Li-ion batteries. \emph{Digital Discovery} \textbf{2025}, \emph{4}, 1544--1559\relax
\mciteBstWouldAddEndPuncttrue
\mciteSetBstMidEndSepPunct{\mcitedefaultmidpunct}
{\mcitedefaultendpunct}{\mcitedefaultseppunct}\relax
\EndOfBibitem
\bibitem[Bergwerf()]{molview_bergwerf}
Bergwerf,~H. MolView: An open-source, web-based web application to make science and education more awesome! \url{https://molview.org}, Accessed: Mar. 23, 2026\relax
\mciteBstWouldAddEndPuncttrue
\mciteSetBstMidEndSepPunct{\mcitedefaultmidpunct}
{\mcitedefaultendpunct}{\mcitedefaultseppunct}\relax
\EndOfBibitem
\bibitem[Hurle and Woolf(1982)Hurle, and Woolf]{packmole_hurle1982self}
Hurle,~R.~L.; Woolf,~L.~A. Self-diffusion in liquid acetonitrile under pressure. \emph{Journal of the Chemical Society, Faraday Transactions 1: Physical Chemistry in Condensed Phases} \textbf{1982}, \emph{78}, 2233--2238\relax
\mciteBstWouldAddEndPuncttrue
\mciteSetBstMidEndSepPunct{\mcitedefaultmidpunct}
{\mcitedefaultendpunct}{\mcitedefaultseppunct}\relax
\EndOfBibitem
\bibitem[{\v{S}}imurka \latin{et~al.}(2018){\v{S}}imurka, {\v{C}}tvrtl{\'\i}k, Toma{\v{s}}t{\'\i}k, Bekta{\c{s}}, Svoboda, and Bange]{vsimurka2018mechanical}
{\v{S}}imurka,~L.; {\v{C}}tvrtl{\'\i}k,~R.; Toma{\v{s}}t{\'\i}k,~J.; Bekta{\c{s}},~G.; Svoboda,~J.; Bange,~K. Mechanical and optical properties of SiO2 thin films deposited on glass. \emph{Chemical Papers} \textbf{2018}, \emph{72}, 2143--2151\relax
\mciteBstWouldAddEndPuncttrue
\mciteSetBstMidEndSepPunct{\mcitedefaultmidpunct}
{\mcitedefaultendpunct}{\mcitedefaultseppunct}\relax
\EndOfBibitem
\bibitem[Dehghani and Soleimani(2021)Dehghani, and Soleimani]{dehghani2021effect}
Dehghani,~P.; Soleimani,~F. Effect of cristobalite content on physical, dielectric constant, and bending strength of fused silica ceramics formed by slip casting method. \emph{Advanced Ceramics Progress} \textbf{2021}, \emph{7}, 16--22\relax
\mciteBstWouldAddEndPuncttrue
\mciteSetBstMidEndSepPunct{\mcitedefaultmidpunct}
{\mcitedefaultendpunct}{\mcitedefaultseppunct}\relax
\EndOfBibitem
\bibitem[An \latin{et~al.}(2026)An, Oh, Lee, Ko, Oh, Hong, and Han]{chfetch_An_2026}
An,~H.; Oh,~S.; Lee,~D.; Ko,~J.-h.; Oh,~D.; Hong,~C.; Han,~S. Etching-to-deposition transition in SiO$_2$/Si$_3$N$_4$ using CH$_x$F$_y$ ion-based plasma etching: An atomistic study with neural network potentials. \emph{Applied Surface Science} \textbf{2026}, \emph{716}, 164601\relax
\mciteBstWouldAddEndPuncttrue
\mciteSetBstMidEndSepPunct{\mcitedefaultmidpunct}
{\mcitedefaultendpunct}{\mcitedefaultseppunct}\relax
\EndOfBibitem
\bibitem[Tully(1977)]{nion_tully1977neutralization}
Tully,~J.~C. Neutralization of ions at surfaces. \emph{Physical Review B} \textbf{1977}, \emph{16}, 4324\relax
\mciteBstWouldAddEndPuncttrue
\mciteSetBstMidEndSepPunct{\mcitedefaultmidpunct}
{\mcitedefaultendpunct}{\mcitedefaultseppunct}\relax
\EndOfBibitem
\bibitem[Pretzer and Hagstrum(1966)Pretzer, and Hagstrum]{nion_pretzer1966ion}
Pretzer,~D.; Hagstrum,~H. {Ion neutralization studies of the (111), ($\bar{1}\bar{1}\bar{1}$) and (110) surfaces of GaAs}. \emph{Surface Science} \textbf{1966}, \emph{4}, 265--285\relax
\mciteBstWouldAddEndPuncttrue
\mciteSetBstMidEndSepPunct{\mcitedefaultmidpunct}
{\mcitedefaultendpunct}{\mcitedefaultseppunct}\relax
\EndOfBibitem
\bibitem[Allen and Tildesley(2017)Allen, and Tildesley]{langevin_allen2017computer}
Allen,~M.~P.; Tildesley,~D.~J. \emph{Computer simulation of liquids}; Oxford University Press, 2017\relax
\mciteBstWouldAddEndPuncttrue
\mciteSetBstMidEndSepPunct{\mcitedefaultmidpunct}
{\mcitedefaultendpunct}{\mcitedefaultseppunct}\relax
\EndOfBibitem
\bibitem[Mogab \latin{et~al.}(1978)Mogab, Adams, and Flamm]{mogab1978plasma}
Mogab,~C.; Adams,~A.; Flamm,~D. {Plasma etching of Si and SiO$_2$—The effect of oxygen additions to CF$_4$ plasmas}. \emph{Journal of applied physics} \textbf{1978}, \emph{49}, 3796--3803\relax
\mciteBstWouldAddEndPuncttrue
\mciteSetBstMidEndSepPunct{\mcitedefaultmidpunct}
{\mcitedefaultendpunct}{\mcitedefaultseppunct}\relax
\EndOfBibitem
\bibitem[Toyoda \latin{et~al.}(2004)Toyoda, Morishima, Fukute, Hori, Murakami, and Sugai]{ionbeam_toyoda2004beam}
Toyoda,~H.; Morishima,~H.; Fukute,~R.; Hori,~Y.; Murakami,~I.; Sugai,~H. {Beam study of the Si and SiO$_2$ etching processes by energetic fluorocarbon ions}. \emph{Journal of applied physics} \textbf{2004}, \emph{95}, 5172--5179\relax
\mciteBstWouldAddEndPuncttrue
\mciteSetBstMidEndSepPunct{\mcitedefaultmidpunct}
{\mcitedefaultendpunct}{\mcitedefaultseppunct}\relax
\EndOfBibitem
\bibitem[Kresse and Joubert(1999)Kresse, and Joubert]{vasp_kresse1999ultrasoft}
Kresse,~G.; Joubert,~D. From ultrasoft pseudopotentials to the projector augmented-wave method. \emph{Physical Review B} \textbf{1999}, \emph{59}, 1758\relax
\mciteBstWouldAddEndPuncttrue
\mciteSetBstMidEndSepPunct{\mcitedefaultmidpunct}
{\mcitedefaultendpunct}{\mcitedefaultseppunct}\relax
\EndOfBibitem
\bibitem[Ong \latin{et~al.}(2013)Ong, Richards, Jain, Hautier, Kocher, Cholia, Gunter, Chevrier, Persson, and Ceder]{pymatgen_ong2013python}
Ong,~S.~P.; Richards,~W.~D.; Jain,~A.; Hautier,~G.; Kocher,~M.; Cholia,~S.; Gunter,~D.; Chevrier,~V.~L.; Persson,~K.~A.; Ceder,~G. Python Materials Genomics (pymatgen): A robust, open-source python library for materials analysis. \emph{Computational Materials Science} \textbf{2013}, \emph{68}, 314--319\relax
\mciteBstWouldAddEndPuncttrue
\mciteSetBstMidEndSepPunct{\mcitedefaultmidpunct}
{\mcitedefaultendpunct}{\mcitedefaultseppunct}\relax
\EndOfBibitem
\bibitem[Batatia \latin{et~al.}(2025)Batatia, Benner, Chiang, Elena, Kov{\'a}cs, Riebesell, Advincula, Asta, Avaylon, Baldwin, and {et al.}]{mace_batatia2025foundation}
Batatia,~I.; Benner,~P.; Chiang,~Y.; Elena,~A.~M.; Kov{\'a}cs,~D.~P.; Riebesell,~J.; Advincula,~X.~R.; Asta,~M.; Avaylon,~M.; Baldwin,~W.~J.; {et al.} A foundation model for atomistic materials chemistry. \emph{The Journal of Chemical Physics} \textbf{2025}, \emph{163}\relax
\mciteBstWouldAddEndPuncttrue
\mciteSetBstMidEndSepPunct{\mcitedefaultmidpunct}
{\mcitedefaultendpunct}{\mcitedefaultseppunct}\relax
\EndOfBibitem
\bibitem[Riebesell \latin{et~al.}(2025)Riebesell, Goodall, Benner, Chiang, Deng, Ceder, Asta, Lee, Jain, and Persson]{matbench_riebesell2025framework}
Riebesell,~J.; Goodall,~R.~E.; Benner,~P.; Chiang,~Y.; Deng,~B.; Ceder,~G.; Asta,~M.; Lee,~A.~A.; Jain,~A.; Persson,~K.~A. A framework to evaluate machine learning crystal stability predictions. \emph{Nature Machine Intelligence} \textbf{2025}, \emph{7}, 836--847\relax
\mciteBstWouldAddEndPuncttrue
\mciteSetBstMidEndSepPunct{\mcitedefaultmidpunct}
{\mcitedefaultendpunct}{\mcitedefaultseppunct}\relax
\EndOfBibitem
\bibitem[Zheng \latin{et~al.}(2020)Zheng, Li, Tran, Chen, Horton, Winston, Persson, and Ong]{b_gb_zheng2020grain}
Zheng,~H.; Li,~X.-G.; Tran,~R.; Chen,~C.; Horton,~M.; Winston,~D.; Persson,~K.~A.; Ong,~S.~P. Grain boundary properties of elemental metals. \emph{Acta Materialia} \textbf{2020}, \emph{186}, 40--49\relax
\mciteBstWouldAddEndPuncttrue
\mciteSetBstMidEndSepPunct{\mcitedefaultmidpunct}
{\mcitedefaultendpunct}{\mcitedefaultseppunct}\relax
\EndOfBibitem
\bibitem[Becquart \latin{et~al.}(2007)Becquart, Raulot, Bencteux, Domain, Perez, Garruchet, and Nguyen]{b_df_becquart2007atomistic}
Becquart,~C.; Raulot,~J.-M.; Bencteux,~G.; Domain,~C.; Perez,~M.; Garruchet,~S.; Nguyen,~H. Atomistic modeling of an Fe system with a small concentration of C. \emph{Computational materials science} \textbf{2007}, \emph{40}, 119--129\relax
\mciteBstWouldAddEndPuncttrue
\mciteSetBstMidEndSepPunct{\mcitedefaultmidpunct}
{\mcitedefaultendpunct}{\mcitedefaultseppunct}\relax
\EndOfBibitem
\bibitem[Olsson \latin{et~al.}(2010)Olsson, Klaver, and Domain]{b_df_olsson2010ab}
Olsson,~P.; Klaver,~T.; Domain,~C. Ab initio study of solute transition-metal interactions with point defects in bcc Fe. \emph{Physical Review B—Condensed Matter and Materials Physics} \textbf{2010}, \emph{81}, 054102\relax
\mciteBstWouldAddEndPuncttrue
\mciteSetBstMidEndSepPunct{\mcitedefaultmidpunct}
{\mcitedefaultendpunct}{\mcitedefaultseppunct}\relax
\EndOfBibitem
\bibitem[Lahey \latin{et~al.}(2020)Lahey, Thien~Phuc, and Rowley]{b_tor_lahey2020benchmarking}
Lahey,~S.-L.~J.; Thien~Phuc,~T.~N.; Rowley,~C.~N. Benchmarking force field and the ani neural network potentials for the torsional potential energy surface of biaryl drug fragments. \emph{Journal of Chemical Information and Modeling} \textbf{2020}, \emph{60}, 6258--6268\relax
\mciteBstWouldAddEndPuncttrue
\mciteSetBstMidEndSepPunct{\mcitedefaultmidpunct}
{\mcitedefaultendpunct}{\mcitedefaultseppunct}\relax
\EndOfBibitem
\bibitem[Rai \latin{et~al.}(2022)Rai, Sresht, Yang, Unwalla, Tu, Mathiowetz, and Bakken]{b_tor_rai2022torsionnet}
Rai,~B.~K.; Sresht,~V.; Yang,~Q.; Unwalla,~R.; Tu,~M.; Mathiowetz,~A.~M.; Bakken,~G.~A. Torsionnet: A deep neural network to rapidly predict small-molecule torsional energy profiles with the accuracy of quantum mechanics. \emph{Journal of Chemical Information and Modeling} \textbf{2022}, \emph{62}, 785--800\relax
\mciteBstWouldAddEndPuncttrue
\mciteSetBstMidEndSepPunct{\mcitedefaultmidpunct}
{\mcitedefaultendpunct}{\mcitedefaultseppunct}\relax
\EndOfBibitem
\bibitem[Dohm \latin{et~al.}(2018)Dohm, Hansen, Steinmetz, Grimme, and Checinski]{b_om_dohm2018comprehensive}
Dohm,~S.; Hansen,~A.; Steinmetz,~M.; Grimme,~S.; Checinski,~M.~P. Comprehensive thermochemical benchmark set of realistic closed-shell metal organic reactions. \emph{Journal of Chemical Theory and Computation} \textbf{2018}, \emph{14}, 2596--2608\relax
\mciteBstWouldAddEndPuncttrue
\mciteSetBstMidEndSepPunct{\mcitedefaultmidpunct}
{\mcitedefaultendpunct}{\mcitedefaultseppunct}\relax
\EndOfBibitem
\bibitem[Gusev(2013)]{b_om_gusev2013assessing}
Gusev,~D.~G. Assessing the accuracy of M06-L organometallic thermochemistry. \emph{Organometallics} \textbf{2013}, \emph{32}, 4239--4243\relax
\mciteBstWouldAddEndPuncttrue
\mciteSetBstMidEndSepPunct{\mcitedefaultmidpunct}
{\mcitedefaultendpunct}{\mcitedefaultseppunct}\relax
\EndOfBibitem
\bibitem[Reilly and Tkatchenko(2013)Reilly, and Tkatchenko]{b_mc_reilly2013understanding}
Reilly,~A.~M.; Tkatchenko,~A. Understanding the role of vibrations, exact exchange, and many-body van der Waals interactions in the cohesive properties of molecular crystals. \emph{The Journal of Chemical Physics} \textbf{2013}, \emph{139}\relax
\mciteBstWouldAddEndPuncttrue
\mciteSetBstMidEndSepPunct{\mcitedefaultmidpunct}
{\mcitedefaultendpunct}{\mcitedefaultseppunct}\relax
\EndOfBibitem
\bibitem[Moellmann and Grimme(2014)Moellmann, and Grimme]{b_mc_moellmann2014dft}
Moellmann,~J.; Grimme,~S. DFT-D3 study of some molecular crystals. \emph{The Journal of Physical Chemistry C} \textbf{2014}, \emph{118}, 7615--7621\relax
\mciteBstWouldAddEndPuncttrue
\mciteSetBstMidEndSepPunct{\mcitedefaultmidpunct}
{\mcitedefaultendpunct}{\mcitedefaultseppunct}\relax
\EndOfBibitem
\bibitem[Zhugayevych \latin{et~al.}(2023)Zhugayevych, Sun, van~der Heide, Lien-Medrano, Frauenheim, and Tretiak]{b_mc_zhugayevych2023benchmark}
Zhugayevych,~A.; Sun,~W.; van~der Heide,~T.; Lien-Medrano,~C.~R.; Frauenheim,~T.; Tretiak,~S. Benchmark data set of crystalline organic semiconductors. \emph{Journal of Chemical Theory and Computation} \textbf{2023}, \emph{19}, 8481--8490\relax
\mciteBstWouldAddEndPuncttrue
\mciteSetBstMidEndSepPunct{\mcitedefaultmidpunct}
{\mcitedefaultendpunct}{\mcitedefaultseppunct}\relax
\EndOfBibitem
\bibitem[Kim \latin{et~al.}(2017)Kim, Huan, Krishnan, and Ramprasad]{b_pv_kim2017hybrid}
Kim,~C.; Huan,~T.~D.; Krishnan,~S.; Ramprasad,~R. A hybrid organic-inorganic perovskite dataset. \emph{Scientific Data} \textbf{2017}, \emph{4}, 1--11\relax
\mciteBstWouldAddEndPuncttrue
\mciteSetBstMidEndSepPunct{\mcitedefaultmidpunct}
{\mcitedefaultendpunct}{\mcitedefaultseppunct}\relax
\EndOfBibitem
\bibitem[Kim \latin{et~al.}(2026)Kim, Kim, Hahm, Kwon, Park, Hong, and Han]{b_asd_kim2026computational}
Kim,~G.; Kim,~P.-h.; Hahm,~S.~G.; Kwon,~M.; Park,~B.; Hong,~C.; Han,~S. A computational study for screening high-selectivity inhibitors in area-selective atomic layer deposition on amorphous surfaces. \emph{Applied Surface Science} \textbf{2026}, \emph{730}, 166294\relax
\mciteBstWouldAddEndPuncttrue
\mciteSetBstMidEndSepPunct{\mcitedefaultmidpunct}
{\mcitedefaultendpunct}{\mcitedefaultseppunct}\relax
\EndOfBibitem
\bibitem[Mallikarjun~Sharada \latin{et~al.}(2019)Mallikarjun~Sharada, Karlsson, Maimaiti, Voss, and Bligaard]{b_ads41_mallikarjun2019adsorption}
Mallikarjun~Sharada,~S.; Karlsson,~R.~K.; Maimaiti,~Y.; Voss,~J.; Bligaard,~T. Adsorption on transition metal surfaces: Transferability and accuracy of DFT using the ADS41 dataset. \emph{Physical Review B} \textbf{2019}, \emph{100}, 035439\relax
\mciteBstWouldAddEndPuncttrue
\mciteSetBstMidEndSepPunct{\mcitedefaultmidpunct}
{\mcitedefaultendpunct}{\mcitedefaultseppunct}\relax
\EndOfBibitem
\bibitem[Kra{\ss} \latin{et~al.}(2025)Kra{\ss}, Huang, and Moosavi]{b_mof_krass2025mofsimbench}
Kra{\ss},~H.; Huang,~J.; Moosavi,~S.~M. MOFSimBench: evaluating universal machine learning interatomic potentials in metal-organic framework molecular modeling. \emph{NPJ Computational Materials} \textbf{2025}, \emph{12}, 4\relax
\mciteBstWouldAddEndPuncttrue
\mciteSetBstMidEndSepPunct{\mcitedefaultmidpunct}
{\mcitedefaultendpunct}{\mcitedefaultseppunct}\relax
\EndOfBibitem
\bibitem[Moosavi \latin{et~al.}(2022)Moosavi, Novotny, Ongari, Moubarak, Asgari, Kadioglu, Charalambous, Ortega-Guerrero, Farmahini, Sarkisov, and {et al.}]{b_mof_moosavi2022data}
Moosavi,~S.~M.; Novotny,~B.~{\'A}.; Ongari,~D.; Moubarak,~E.; Asgari,~M.; Kadioglu,~{\"O}.; Charalambous,~C.; Ortega-Guerrero,~A.; Farmahini,~A.~H.; Sarkisov,~L.; {et al.} A data-science approach to predict the heat capacity of nanoporous materials. \emph{Nature Materials} \textbf{2022}, \emph{21}, 1419--1425\relax
\mciteBstWouldAddEndPuncttrue
\mciteSetBstMidEndSepPunct{\mcitedefaultmidpunct}
{\mcitedefaultendpunct}{\mcitedefaultseppunct}\relax
\EndOfBibitem
\bibitem[Sriram \latin{et~al.}(2025)Sriram, Brabson, Yu, Choi, Abdelmaqsoud, Moubarak, de~Haan, Löwe, Brehmer, Kitchin, Welling, Zitnick, Ulissi, Medford, and Sholl]{b_mof_odac25_sriram2025open}
Sriram,~A.; Brabson,~L.~M.; Yu,~X.; Choi,~S.; Abdelmaqsoud,~K.; Moubarak,~E.; de~Haan,~P.; Löwe,~S.; Brehmer,~J.; Kitchin,~J.~R.; Welling,~M.; Zitnick,~C.~L.; Ulissi,~Z.; Medford,~A.~J.; Sholl,~D.~S. The Open DAC 2025 Dataset for Sorbent Discovery in Direct Air Capture. 2025; \url{https://arxiv.org/abs/2508.03162}\relax
\mciteBstWouldAddEndPuncttrue
\mciteSetBstMidEndSepPunct{\mcitedefaultmidpunct}
{\mcitedefaultendpunct}{\mcitedefaultseppunct}\relax
\EndOfBibitem
\bibitem[Lim \latin{et~al.}(2025)Lim, Park, Walsh, and Kim]{b_mof_golddac_lim2025accelerating}
Lim,~Y.; Park,~H.; Walsh,~A.; Kim,~J. Accelerating CO2 direct air capture screening for metal-organic frameworks with a transferable machine learning force field. \emph{Matter} \textbf{2025}, \emph{8}\relax
\mciteBstWouldAddEndPuncttrue
\mciteSetBstMidEndSepPunct{\mcitedefaultmidpunct}
{\mcitedefaultendpunct}{\mcitedefaultseppunct}\relax
\EndOfBibitem
\bibitem[Brabson \latin{et~al.}(2025)Brabson, Medford, and Sholl]{b_mof_brabson2025comparing}
Brabson,~L.~M.; Medford,~A.~J.; Sholl,~D.~S. Comparing classical and machine learning force fields for modeling deformation of metal--organic frameworks relevant for direct air capture. \emph{The Journal of Physical Chemistry C} \textbf{2025}, \emph{129}, 16811--16825\relax
\mciteBstWouldAddEndPuncttrue
\mciteSetBstMidEndSepPunct{\mcitedefaultmidpunct}
{\mcitedefaultendpunct}{\mcitedefaultseppunct}\relax
\EndOfBibitem
\bibitem[Egger \latin{et~al.}(2016)Egger, Rappe, and Kronik]{hybprv_egger2016hybrid}
Egger,~D.~A.; Rappe,~A.~M.; Kronik,~L. Hybrid organic--inorganic perovskites on the move. \emph{Accounts of Chemical Research} \textbf{2016}, \emph{49}, 573--581\relax
\mciteBstWouldAddEndPuncttrue
\mciteSetBstMidEndSepPunct{\mcitedefaultmidpunct}
{\mcitedefaultendpunct}{\mcitedefaultseppunct}\relax
\EndOfBibitem
\bibitem[Moroni \latin{et~al.}(2024)Moroni, Coccia, and Malavasi]{hybprv_moroni2024chiral}
Moroni,~M.; Coccia,~C.; Malavasi,~L. Chiral 2D and quasi-2D hybrid organic inorganic perovskites: from fundamentals to applications. \emph{Chemical Communications} \textbf{2024}, \emph{60}, 9310--9327\relax
\mciteBstWouldAddEndPuncttrue
\mciteSetBstMidEndSepPunct{\mcitedefaultmidpunct}
{\mcitedefaultendpunct}{\mcitedefaultseppunct}\relax
\EndOfBibitem
\bibitem[Berry \latin{et~al.}(2015)Berry, Buonassisi, Egger, Hodes, Kronik, Loo, Lubomirsky, Marder, Mastai, Miller, and {et al.}]{hybprv_berry2015hybrid}
Berry,~J.; Buonassisi,~T.; Egger,~D.~A.; Hodes,~G.; Kronik,~L.; Loo,~Y.-L.; Lubomirsky,~I.; Marder,~S.~R.; Mastai,~Y.; Miller,~J.~S.; {et al.} Hybrid organic--inorganic perovskites (HOIPs): opportunities and challenges. \emph{Advanced Materials} \textbf{2015}, \emph{27}, 5102--5112\relax
\mciteBstWouldAddEndPuncttrue
\mciteSetBstMidEndSepPunct{\mcitedefaultmidpunct}
{\mcitedefaultendpunct}{\mcitedefaultseppunct}\relax
\EndOfBibitem
\bibitem[Ding \latin{et~al.}(2019)Ding, Flaig, Jiang, and Yaghi]{mof_ding2019carbon}
Ding,~M.; Flaig,~R.~W.; Jiang,~H.-L.; Yaghi,~O.~M. Carbon capture and conversion using metal--organic frameworks and MOF-based materials. \emph{Chemical Society Reviews} \textbf{2019}, \emph{48}, 2783--2828\relax
\mciteBstWouldAddEndPuncttrue
\mciteSetBstMidEndSepPunct{\mcitedefaultmidpunct}
{\mcitedefaultendpunct}{\mcitedefaultseppunct}\relax
\EndOfBibitem
\bibitem[Simmons \latin{et~al.}(2011)Simmons, Wu, Zhou, and Yildirim]{mof_simmons2011carbon}
Simmons,~J.~M.; Wu,~H.; Zhou,~W.; Yildirim,~T. Carbon capture in metal--organic frameworks—a comparative study. \emph{Energy \& Environmental Science} \textbf{2011}, \emph{4}, 2177--2185\relax
\mciteBstWouldAddEndPuncttrue
\mciteSetBstMidEndSepPunct{\mcitedefaultmidpunct}
{\mcitedefaultendpunct}{\mcitedefaultseppunct}\relax
\EndOfBibitem
\bibitem[N{\o}rskov \latin{et~al.}(2009)N{\o}rskov, Bligaard, Rossmeisl, and Christensen]{msur_norskov2009towards}
N{\o}rskov,~J.~K.; Bligaard,~T.; Rossmeisl,~J.; Christensen,~C.~H. Towards the computational design of solid catalysts. \emph{Nature Chemistry} \textbf{2009}, \emph{1}, 37--46\relax
\mciteBstWouldAddEndPuncttrue
\mciteSetBstMidEndSepPunct{\mcitedefaultmidpunct}
{\mcitedefaultendpunct}{\mcitedefaultseppunct}\relax
\EndOfBibitem
\bibitem[Trepte and Voss(2022)Trepte, and Voss]{b_ads41_trepte2022data}
Trepte,~K.; Voss,~J. Data-driven and constrained optimization of semi-local exchange and nonlocal correlation functionals for materials and surface chemistry. \emph{Journal of Computational Chemistry} \textbf{2022}, \emph{43}, 1104--1112\relax
\mciteBstWouldAddEndPuncttrue
\mciteSetBstMidEndSepPunct{\mcitedefaultmidpunct}
{\mcitedefaultendpunct}{\mcitedefaultseppunct}\relax
\EndOfBibitem
\bibitem[Lee \latin{et~al.}(2025)Lee, Rothman, Shearer, and Bent]{asd_lee2025molecular}
Lee,~Y.; Rothman,~A.; Shearer,~A.~B.; Bent,~S.~F. Molecular design in Area-Selective atomic layer deposition: Understanding inhibitors and precursors. \emph{Chemistry of Materials} \textbf{2025}, \emph{37}, 1741--1758\relax
\mciteBstWouldAddEndPuncttrue
\mciteSetBstMidEndSepPunct{\mcitedefaultmidpunct}
{\mcitedefaultendpunct}{\mcitedefaultseppunct}\relax
\EndOfBibitem
\bibitem[Zhao \latin{et~al.}(2020)Zhao, Stalin, Zhao, and Archer]{sse_zhao2020designing}
Zhao,~Q.; Stalin,~S.; Zhao,~C.-Z.; Archer,~L.~A. Designing solid-state electrolytes for safe, energy-dense batteries. \emph{Nature Reviews Materials} \textbf{2020}, \emph{5}, 229--252\relax
\mciteBstWouldAddEndPuncttrue
\mciteSetBstMidEndSepPunct{\mcitedefaultmidpunct}
{\mcitedefaultendpunct}{\mcitedefaultseppunct}\relax
\EndOfBibitem
\bibitem[Deng \latin{et~al.}(2025)Deng, Choi, Zhong, Riebesell, Anand, Li, Jun, Persson, and Ceder]{ftsoft_deng2025systematic}
Deng,~B.; Choi,~Y.; Zhong,~P.; Riebesell,~J.; Anand,~S.; Li,~Z.; Jun,~K.; Persson,~K.~A.; Ceder,~G. Systematic softening in universal machine learning interatomic potentials. \emph{npj Computational Materials} \textbf{2025}, \emph{11}, 9\relax
\mciteBstWouldAddEndPuncttrue
\mciteSetBstMidEndSepPunct{\mcitedefaultmidpunct}
{\mcitedefaultendpunct}{\mcitedefaultseppunct}\relax
\EndOfBibitem
\bibitem[Hänseroth \latin{et~al.}(2026)Hänseroth, Flötotto, Qaisrani, and Dreßler]{ftset_roth2026fine}
Hänseroth,~J.; Flötotto,~A.; Qaisrani,~M.~N.; Dreßler,~C. Fine-Tuning Unifies Foundational Machine-Learned Interatomic Potential Architectures at ab initio Accuracy. \emph{The Journal of Physical Chemistry Letters} \textbf{2026}, \emph{17}, 3152--3162\relax
\mciteBstWouldAddEndPuncttrue
\mciteSetBstMidEndSepPunct{\mcitedefaultmidpunct}
{\mcitedefaultendpunct}{\mcitedefaultseppunct}\relax
\EndOfBibitem
\bibitem[Kaur \latin{et~al.}(2025)Kaur, Della~Pia, Batatia, Advincula, Shi, Lan, Cs{\'a}nyi, Michaelides, and Kapil]{ftset_kaur2025data}
Kaur,~H.; Della~Pia,~F.; Batatia,~I.; Advincula,~X.~R.; Shi,~B.~X.; Lan,~J.; Cs{\'a}nyi,~G.; Michaelides,~A.; Kapil,~V. Data-efficient fine-tuning of foundational models for first-principles quality sublimation enthalpies. \emph{Faraday Discussions} \textbf{2025}, \emph{256}, 120--138\relax
\mciteBstWouldAddEndPuncttrue
\mciteSetBstMidEndSepPunct{\mcitedefaultmidpunct}
{\mcitedefaultendpunct}{\mcitedefaultseppunct}\relax
\EndOfBibitem
\bibitem[Xu(2004)]{liqexp_xu2004nonaqueous}
Xu,~K. Nonaqueous liquid electrolytes for lithium-based rechargeable batteries. \emph{Chemical Reviews} \textbf{2004}, \emph{104}, 4303--4418\relax
\mciteBstWouldAddEndPuncttrue
\mciteSetBstMidEndSepPunct{\mcitedefaultmidpunct}
{\mcitedefaultendpunct}{\mcitedefaultseppunct}\relax
\EndOfBibitem
\bibitem[Gong \latin{et~al.}(2025)Gong, Zhang, Mu, Pu, Wang, Han, Yu, Chen, Zheng, Wang, and {et al.}]{liqexp_gong2025predictive}
Gong,~S.; Zhang,~Y.; Mu,~Z.; Pu,~Z.; Wang,~H.; Han,~X.; Yu,~Z.; Chen,~M.; Zheng,~T.; Wang,~Z.; {et al.} A predictive machine learning force-field framework for liquid electrolyte development. \emph{Nature Machine Intelligence} \textbf{2025}, \emph{7}, 543--552\relax
\mciteBstWouldAddEndPuncttrue
\mciteSetBstMidEndSepPunct{\mcitedefaultmidpunct}
{\mcitedefaultendpunct}{\mcitedefaultseppunct}\relax
\EndOfBibitem
\bibitem[Almasi \latin{et~al.}(2023)Almasi, Hernandez, and Zarrini]{liqexp_almasi2023density}
Almasi,~M.; Hernandez,~A.; Zarrini,~J. Density and viscosity studies of ethylene glycol diethyl ether and 1-alkanol mixtures. \emph{Journal of Chemical \& Engineering Data} \textbf{2023}, \emph{68}, 1911--1920\relax
\mciteBstWouldAddEndPuncttrue
\mciteSetBstMidEndSepPunct{\mcitedefaultmidpunct}
{\mcitedefaultendpunct}{\mcitedefaultseppunct}\relax
\EndOfBibitem
\bibitem[Fran{\c{c}}a \latin{et~al.}(2009)Fran{\c{c}}a, Nieto~de Castro, Lopes, and Nunes]{liqexp_francca2009influence}
Fran{\c{c}}a,~J.~M.; Nieto~de Castro,~C.~A.; Lopes,~M.~M.; Nunes,~V.~M. Influence of thermophysical properties of ionic liquids in chemical process design. \emph{Journal of Chemical \& Engineering Data} \textbf{2009}, \emph{54}, 2569--2575\relax
\mciteBstWouldAddEndPuncttrue
\mciteSetBstMidEndSepPunct{\mcitedefaultmidpunct}
{\mcitedefaultendpunct}{\mcitedefaultseppunct}\relax
\EndOfBibitem
\bibitem[Guard(1999)]{liqexp_guard1999chemical}
Guard,~U.~C. Chemical hazard response information system (CHRIS)-hazardous chemical data. \emph{Commandant Instruction} \textbf{1999}, \emph{16465}, 3--550\relax
\mciteBstWouldAddEndPuncttrue
\mciteSetBstMidEndSepPunct{\mcitedefaultmidpunct}
{\mcitedefaultendpunct}{\mcitedefaultseppunct}\relax
\EndOfBibitem
\bibitem[Hagiyama \latin{et~al.}(2008)Hagiyama, Suzuki, Ohtake, Shimada, Nanbu, Takehara, Ue, and Sasaki]{liqexp_hagiyama2008physical}
Hagiyama,~K.; Suzuki,~K.; Ohtake,~M.; Shimada,~M.; Nanbu,~N.; Takehara,~M.; Ue,~M.; Sasaki,~Y. Physical properties of substituted 1, 3-dioxolan-2-ones. \emph{Chemistry Letters} \textbf{2008}, \emph{37}, 210--211\relax
\mciteBstWouldAddEndPuncttrue
\mciteSetBstMidEndSepPunct{\mcitedefaultmidpunct}
{\mcitedefaultendpunct}{\mcitedefaultseppunct}\relax
\EndOfBibitem
\bibitem[V{\"a}li \latin{et~al.}(2016)V{\"a}li, J{\"a}nes, and Lust]{liqexp_vali2016vinylene}
V{\"a}li,~R.; J{\"a}nes,~A.; Lust,~E. Vinylene carbonate as co-solvent for low-temperature mixed electrolyte based Supercapacitors. \emph{Journal of The Electrochemical Society} \textbf{2016}, \emph{163}, A851--A857\relax
\mciteBstWouldAddEndPuncttrue
\mciteSetBstMidEndSepPunct{\mcitedefaultmidpunct}
{\mcitedefaultendpunct}{\mcitedefaultseppunct}\relax
\EndOfBibitem
\bibitem[Yang \latin{et~al.}(2007)Yang, Peng, Huang, Fan, and Yang]{liqexp_yang2007study}
Yang,~S.-K.; Peng,~S.-J.; Huang,~J.-H.; Fan,~L.-Q.; Yang,~F.-X. A study on densities and excess volumes in the ($\gamma$-butyrolactone+ aromatic hydrocarbon) system at various temperatures. \emph{The Journal of Chemical Thermodynamics} \textbf{2007}, \emph{39}, 773--780\relax
\mciteBstWouldAddEndPuncttrue
\mciteSetBstMidEndSepPunct{\mcitedefaultmidpunct}
{\mcitedefaultendpunct}{\mcitedefaultseppunct}\relax
\EndOfBibitem
\bibitem[Zheng \latin{et~al.}(2008)Zheng, Meng, Wu, and Liu]{liqexp_zheng2008density}
Zheng,~P.; Meng,~X.; Wu,~J.; Liu,~Z. Density and Viscosity Measurements of Dimethoxymethane and 1, 2-Dimethoxyethane from 243 K to 373 K up to 20 MPa. \emph{International Journal of Thermophysics} \textbf{2008}, \emph{29}, 1244--1256\relax
\mciteBstWouldAddEndPuncttrue
\mciteSetBstMidEndSepPunct{\mcitedefaultmidpunct}
{\mcitedefaultendpunct}{\mcitedefaultseppunct}\relax
\EndOfBibitem
\bibitem[Cao \latin{et~al.}(2023)Cao, Bu, Vinet, Cao, Takagi, Hwang, Ghani, and Banerjee]{har_cao2023future}
Cao,~W.; Bu,~H.; Vinet,~M.; Cao,~M.; Takagi,~S.; Hwang,~S.; Ghani,~T.; Banerjee,~K. The future transistors. \emph{Nature} \textbf{2023}, \emph{620}, 501--515\relax
\mciteBstWouldAddEndPuncttrue
\mciteSetBstMidEndSepPunct{\mcitedefaultmidpunct}
{\mcitedefaultendpunct}{\mcitedefaultseppunct}\relax
\EndOfBibitem
\bibitem[Hayashi \latin{et~al.}(1996)Hayashi, Kazuaki~Kurihara, and Makoto~Sekine]{sio2etch_hayashi1996char}
Hayashi,~H.; Kazuaki~Kurihara,~K.~K.; Makoto~Sekine,~M.~S. Characterization of Highly Selective SiO$_2$/Si$_3$N$_4$ Etching of High-Aspect-Ratio Holes. \emph{Japanese Journal of Applied Physics} \textbf{1996}, \emph{35}, 2488\relax
\mciteBstWouldAddEndPuncttrue
\mciteSetBstMidEndSepPunct{\mcitedefaultmidpunct}
{\mcitedefaultendpunct}{\mcitedefaultseppunct}\relax
\EndOfBibitem
\bibitem[Lin \latin{et~al.}(2018)Lin, Li, Engelmann, Bruce, Joseph, Metzler, and Oehrlein]{sio2etch_lin2018achieving}
Lin,~K.-Y.; Li,~C.; Engelmann,~S.; Bruce,~R.~L.; Joseph,~E.~A.; Metzler,~D.; Oehrlein,~G.~S. Achieving ultrahigh etching selectivity of SiO2 over Si3N4 and Si in atomic layer etching by exploiting chemistry of complex hydrofluorocarbon precursors. \emph{Journal of Vacuum Science \& Technology A} \textbf{2018}, \emph{36}\relax
\mciteBstWouldAddEndPuncttrue
\mciteSetBstMidEndSepPunct{\mcitedefaultmidpunct}
{\mcitedefaultendpunct}{\mcitedefaultseppunct}\relax
\EndOfBibitem
\bibitem[Lee \latin{et~al.}(2010)Lee, Oh, Lee, and Sohn]{sio2etch_lee2010ultrahigh}
Lee,~S.; Oh,~J.; Lee,~K.; Sohn,~H. Ultrahigh selective etching of Si3N4 films over SiO2 films for silicon nitride gate spacer etching. \emph{Journal of Vacuum Science \& Technology B} \textbf{2010}, \emph{28}, 131--137\relax
\mciteBstWouldAddEndPuncttrue
\mciteSetBstMidEndSepPunct{\mcitedefaultmidpunct}
{\mcitedefaultendpunct}{\mcitedefaultseppunct}\relax
\EndOfBibitem
\bibitem[Hong \latin{et~al.}(2024)Hong, Oh, An, Kim, Kim, Ko, Sue, Oh, Park, and Han]{hfetch_Hong_2024}
Hong,~C.; Oh,~S.; An,~H.; Kim,~P.-h.; Kim,~Y.; Ko,~J.-h.; Sue,~J.; Oh,~D.; Park,~S.; Han,~S. Atomistic Simulation of HF Etching Process of Amorphous Si3N4 Using Machine Learning Potential. \emph{ACS Applied Materials \& Interfaces} \textbf{2024}, \emph{16}, 48457–48469\relax
\mciteBstWouldAddEndPuncttrue
\mciteSetBstMidEndSepPunct{\mcitedefaultmidpunct}
{\mcitedefaultendpunct}{\mcitedefaultseppunct}\relax
\EndOfBibitem
\bibitem[Steinbr{\"u}chel(1989)]{etch_rel_steinbruchel1989universal}
Steinbr{\"u}chel,~C. Universal energy dependence of physical and ion-enhanced chemical etch yields at low ion energy. \emph{Applied Physics Letters} \textbf{1989}, \emph{55}, 1960--1962\relax
\mciteBstWouldAddEndPuncttrue
\mciteSetBstMidEndSepPunct{\mcitedefaultmidpunct}
{\mcitedefaultendpunct}{\mcitedefaultseppunct}\relax
\EndOfBibitem
\bibitem[Karahashi \latin{et~al.}(2004)Karahashi, Yanai, Ishikawa, Tsuboi, Kurihara, and Nakamura]{ionbeam_karahashi2004etching}
Karahashi,~K.; Yanai,~K.-i.; Ishikawa,~K.; Tsuboi,~H.; Kurihara,~K.; Nakamura,~M. {Etching yield of SiO$_2$ irradiated by F$^+$, CF$_x^+$ ($x= 1, 2, 3$) ion with energies from 250 to 2000 eV}. \emph{Journal of Vacuum Science \& Technology A: Vacuum, Surfaces, and Films} \textbf{2004}, \emph{22}, 1166--1168\relax
\mciteBstWouldAddEndPuncttrue
\mciteSetBstMidEndSepPunct{\mcitedefaultmidpunct}
{\mcitedefaultendpunct}{\mcitedefaultseppunct}\relax
\EndOfBibitem
\bibitem[Shibano \latin{et~al.}(1993)Shibano, Fujiwara, Hirayama, Nagata, and Demizu]{ionbeam_shibano1993etching}
Shibano,~T.; Fujiwara,~N.; Hirayama,~M.; Nagata,~H.; Demizu,~K. {Etching yields of SiO$_2$ by low energy CF$_x^+$ and F$^+$ ions}. \emph{Applied Physics Letters} \textbf{1993}, \emph{63}, 2336--2338\relax
\mciteBstWouldAddEndPuncttrue
\mciteSetBstMidEndSepPunct{\mcitedefaultmidpunct}
{\mcitedefaultendpunct}{\mcitedefaultseppunct}\relax
\EndOfBibitem
\bibitem[Yamaguchi \latin{et~al.}(2000)Yamaguchi, Sasaki, and Kadota]{ionbeam_yamaguchi2000etching}
Yamaguchi,~T.; Sasaki,~K.; Kadota,~K. {Etching Efficiency for Si and SiO$_2$ by CF$^+_x$, F$^+$, and C$^+$ Ion Beams Extracted from CF$_4$ Plasmas}. \emph{Plasma Chemistry and Plasma Processing} \textbf{2000}, \emph{20}, 145--157\relax
\mciteBstWouldAddEndPuncttrue
\mciteSetBstMidEndSepPunct{\mcitedefaultmidpunct}
{\mcitedefaultendpunct}{\mcitedefaultseppunct}\relax
\EndOfBibitem
\bibitem[Geiger \latin{et~al.}(2024)Geiger, Kucukbenli, Zandstein, and Tretina]{cueq_geiger2024}
Geiger,~M.; Kucukbenli,~E.; Zandstein,~B.; Tretina,~K. Accelerate drug and material discovery with new math library {NVIDIA} cuEquivariance. 2024; \url{https://developer.nvidia.com/blog/accelerate-drug-and-material-discovery-with-new-math-library-nvidia-cuequivariance/}\relax
\mciteBstWouldAddEndPuncttrue
\mciteSetBstMidEndSepPunct{\mcitedefaultmidpunct}
{\mcitedefaultendpunct}{\mcitedefaultseppunct}\relax
\EndOfBibitem
\end{mcitethebibliography}

\end{document}